\documentclass[rmp,twocolumn,floatfix,showpacs]{revtex4}

\usepackage{graphicx}
\usepackage{bm}
\usepackage{url}

\begin{document}

\title{Colloquium: Statistical mechanics of money, wealth, and income}

\author{Victor M.~Yakovenko}

\affiliation{Department of Physics, University of
  Maryland, College Park, Maryland 20742-4111, USA}

\author{J. Barkley Rosser, Jr.}

\affiliation{Department of Economics, James Madison University,
Harrisonburg, Virginia 22807, USA}

\date{24 December 2009}


\begin{abstract}
This Colloquium reviews statistical models for money, wealth, and income
distributions developed in the econophysics literature since the late
1990s.  By analogy with the Boltzmann-Gibbs distribution of energy in
physics, it is shown that the probability distribution of money is
exponential for certain classes of models with interacting economic
agents.  Alternative scenarios are also reviewed.  Data analysis of
the empirical distributions of wealth and income reveals a two-class
distribution.  The majority of the population belongs to the lower
class, characterized by the exponential (``thermal'') distribution,
whereas a small fraction of the population in the upper class is
characterized by the power-law (``superthermal'') distribution.  The
lower part is very stable, stationary in time, whereas the upper part
is highly dynamical and out of equilibrium. \\

\textsf{\normalsize ``Money, it's a gas.'' Pink Floyd, 
\textit{Dark Side of the Moon}}
\end{abstract}


\maketitle

\tableofcontents

\section{Historical Introduction}
\label{Sec:History}

This Colloquium article is based on the lectures that one of us (V.M.Y.)
has frequently given during the last nine years, when econophysics
became a popular subject.  Econophysics is a new interdisciplinary
research field applying methods of statistical physics to problems in
economics and finance.  The term ``econophysics'' was first introduced
by the theoretical physicist Eugene Stanley in 1995 at the conference
\textit{Dynamics of Complex Systems}, which was held in Kolkata as a 
satellite meeting to the STATPHYS-19 conference in China
\cite{Chakrabarti-history,Carbone-2007}.  The term appeared first 
by \textcite{Stanley-1996} in the
proceedings of the Kolkata conference.  The paper presented a
manifesto of the new field, arguing that ``behavior of large numbers
of humans (as measured, e.g., by economic indices) might conform to
analogs of the scaling laws that have proved useful in describing
systems composed of large numbers of inanimate objects''
\cite{Stanley-1996}.  Soon the first econophysics conferences were
organized: \textit{International Workshop on Econophysics}, Budapest,
1997 and \textit{International Workshop on Econophysics and
  Statistical Finance}, Palermo, 1998 \cite{Carbone-2007}, and the
book \textit{An Introduction to Econophysics} by
\textcite{Stanley-book} was published.

The term econophysics was introduced by analogy with similar
terms, such as astrophysics, geophysics, and biophysics, which
describe applications of physics to different fields.  Particularly
important is the parallel with biophysics, which studies living
organisms, but they still obey the laws of physics.  Econophysics does 
not literally apply the laws of
physics, such as Newton's laws or quantum mechanics, to humans.  It 
uses mathematical methods developed in statistical physics to
study statistical properties of complex economic systems consisting of
a large number of humans.  As such, it may be considered as a branch of
applied theory of probabilities.  However, statistical physics is
distinctly different from mathematical statistics in its focus,
methods, and results.

Originating from physics as a quantitative science, econophysics
emphasizes quantitative analysis of large amounts of economic and
financial data, which became increasingly available with the 
introduction of computers and the Internet.  Econophysics distances
itself from the verbose, narrative, and ideological style of political
economy and is closer to econometrics in its focus.  Studying
mathematical models of a large number of interacting economic agents,
econophysics has much common ground with the agent-based modeling and
simulation.  Correspondingly, it distances itself from the
representative-agent approach of traditional economics, which, by
definition, ignores statistical and heterogeneous aspects of the
economy.

Another direction related to econophysics has been advocated by the
theoretical physicist Serge Galam since early 1980 under the name of
sociophysics \cite{Galam-history}, with the first appearance of the
term by \textcite{Galam-1982}.  It echoes the term ``physique
sociale'' proposed in the nineteenth century by Auguste Comte, the
founder of sociology.  Unlike econophysics, the term ``sociophysics''
did not catch on when first introduced, but it is coming back with the
popularity of econophysics and active support from some physicists
\cite{Stauffer-history,Schweitzer-2003,Weidlich-2000}.  While the
principles of both fields have much in common, econophysics focuses on
the narrower subject of economic behavior of humans, where more
quantitative data is available, whereas sociophysics studies a broader
range of social issues.  The boundary between econophysics and
sociophysics is not sharp, and the two fields enjoy a good rapport
\cite{Econo-Socio-book}.

Historically, statistical mechanics was developed in the second half
of the nineteenth century by James Clerk Maxwell, Ludwig Boltzmann,
and Josiah Willard Gibbs.  These physicists believed in the existence
of atoms and developed mathematical methods for describing their
statistical properties.  There are interesting connections between the
development of statistical physics and statistics of social phenomena,
which were recently highlighted by the science journalist Philip
\textcite{Ball-2002,Ball-book}.

Collection and study of ``social numbers'', such as the rates of
death, birth, and marriage, has been growing progressively since the
seventeenth century \cite[Ch.~3]{Ball-book}.  The term ``statistics''
was introduced in the eighteenth century to denote these studies
dealing with the civil ``states'', and its practitioners were called
``statists''.  Popularization of social statistics in the nineteenth
century is particularly accredited to the Belgian astronomer Adolphe
Quetelet.  Before the 1850s, statistics was considered an empirical
arm of political economy, but then it started to transform into a
general method of quantitative analysis suitable for all disciplines.
It stimulated physicists to develop statistical mechanics in the
second half of the nineteenth century.

Rudolf Clausius started development of the kinetic theory of gases,
but it was James Clerk Maxwell who made a decisive step of deriving
the probability distribution of velocities of molecules in a gas.
Historical studies show \cite[Ch.~3]{Ball-book} that, in developing
statistical mechanics, Maxwell was strongly influenced and encouraged
by the widespread popularity of social statistics at the time
\cite{Gillispie-1963}.\footnote{V.M.Y.\ is grateful to Stephen G.~Brush
  for this reference.}  This approach was further developed by Ludwig
Boltzmann, who was very explicit about its origins
\cite[p.~69]{Ball-book}:
\begin{quote}
  ``The molecules are like individuals, \ldots\ and the properties of
  gases only remain unaltered, because the number of these molecules,
  which on the average have a given state, is constant.''
\end{quote}
In his book \textit{Popul\"are Schrifen}, \textcite{Boltzmann-1905}
praises Josiah Willard Gibbs for systematic development of statistical
mechanics.  Then, Boltzmann says:\footnote{Cited from
  \textcite{Boltzmann-2006}. V.M.Y.\ is grateful to Michael E.~Fisher for
  this quote.}
\begin{quote}
  ``This opens a broad perspective, if we do not only think of
  mechanical objects.  Let's consider to apply this method to the
  statistics of living beings, society, sociology and so forth.''
\end{quote}

It is worth noting that many now-famous economists were originally
educated in physics and engineering.  Vilfredo Pareto earned a degree
in mathematical sciences and a doctorate in engineering.  Working as a
civil engineer, he collected statistics demonstrating that
distributions of income and wealth in a society follow a power law
\cite{Pareto-book}.  He later became a professor of economics at
Lausanne, where he replaced L\'eon Walras, also an engineer by
education.  The influential American economist Irving Fisher was a
student of Gibbs.  However, most of the mathematical apparatus
transferred to economics from physics was that of Newtonian mechanics
and classical thermodynamics \cite{Mirowski-book,Smith-2008}.  It
culminated in the neoclassical concept of mechanistic equilibrium
where the ``forces'' of supply and demand balance each other.  The
more general concept of statistical equilibrium largely eluded
mainstream economics.

With time, both physics and economics became more formal and rigid in
their specializations, and the social origin of statistical physics
was forgotten.  The situation is well summarized by Philip Ball
\cite[p.~69]{Ball-book}:
\begin{quote}
  ``Today physicists regard the application of statistical mechanics
  to social phenomena as a new and risky venture.  Few, it seems,
  recall how the process originated the other way around, in the days
  when physical science and social science were the twin siblings of a
  mechanistic philosophy and when it was not in the least disreputable
  to invoke the habits of people to explain the habits of inanimate
  particles.''
\end{quote}

Some physicists and economists attempted to connect the two
disciplines during the twentieth century.  Frederick
\textcite{Soddy-book}, the Nobel Prize winner in chemistry for his
work on radioactivity, published the book \textit{Wealth, Virtual
Wealth and Debt}, where he argued that the real wealth is derived
from the energy use in transforming raw materials into goods and
services, and not from monetary transactions.  He also warned about
dangers of excessive debt and related ``virtual wealth'', thus
anticipating the Great Depression.  His ideas were largely ignored at
the time, but resonate today \cite{Defilla-2007}.  The theoretical
physicist Ettore \textcite{Majorana-1942} argued in favor of applying
the laws of statistical physics to social phenomena in a paper
published after his mysterious disappearance.  The statistical
physicist Elliott Montroll co-authored the book \textit{Introduction
to Quantitative Aspects of Social Phenomena} \cite{Montroll-book}.
Several economists
\cite{Follmer-1974,Blume-1993,Foley-1994,Durlauf-1997} applied
statistical physics to economic problems.  The mathematicians
\textcite{Farjoun-1983} argued that many paradoxes in classical
political economy can be resolved if one adopts a probabilistic
approach.  An early attempt to bring together the leading theoretical
physicists and economists at the Santa Fe Institute was not entirely
successful \cite{SantaFe-1988}.  However, by the late 1990s, the
attempts to apply statistical physics to social phenomena finally
coalesced into the robust movements of econophysics and sociophysics.

Current standing of econophysics within the physics and economics
communities is mixed.  Although an entry on econophysics has appeared
in the \textit{New Palgrave Dictionary of Economics} 
\cite{Rosser-2008a}, it is fair to say that econophysics has not been
accepted yet by mainstream economics.  Nevertheless, a number of
open-minded, nontraditional economists have joined this movement, and
the number is growing.  Under these circumstances, econophysicists
have most of their papers published in physics journals.  The journal
\textit{Physica A: Statistical Mechanics and its Applications} has
emerged as the leader in econophysics publications and has even
attracted submissions from some \textit{bona fide} economists.
Gradually, reputable economics journals are also starting to publish
econophysics papers \cite{Lux-2002,Gabaix-2006,Wyart-2007}.  The
mainstream physics community is generally sympathetic to econophysics,
although it is not uncommon for econophysics papers to be rejected by
\textit{Physical Review Letters} on the grounds that ``it is not
physics''.  There is a PACS number for econophysics, and
\textit{Physical Review E} has published many papers on this subject.
There are regular conferences on econophysics, such as
\textit{Applications of Physics in Financial Analysis} (sponsored by
the European Physical Society), \textit{Nikkei Econophysics
  Symposium}, \textit{Econophysics Colloquium}, and
\textit{Econophys-Kolkata} \cite{Kolkata-2005,Chakrabarti-history}.
Econophysics sessions are included in the annual meetings of physical
societies and statistical physics conferences.  The overlap with
economists is the strongest in the field of agent-based simulation.
Not surprisingly, the conference series WEHIA/ESHIA, which deals with
heterogeneous interacting agents, regularly includes sessions on
econophysics.  More information can be found in the reviews by
\textcite{Farmer-2005,Samanidou-2007} and on the Web portal
Econophysics Forum \url{http://www.unifr.ch/econophysics/}.

\section{Statistical Mechanics of Money Distribution}
\label{Sec:money}

When modern econophysics started in the middle of 1990s, its attention
was primarily focused on analysis of financial markets.  Soon after,
another direction, closer to economics than finance, has emerged.  It
studies the probability distributions of money, wealth, and income in
a society and overlaps with the long-standing line of research in
economics studying inequality in a society.\footnote{See, e.g.,\ 
\textcite{Pareto-book,Gibrat-1931,Kalecki-1945,Champernowne-1953,Kakwani-book,Champernowne-1998,Atkinson-2000,Atkinson-2007,Piketty-2003}.}
Many papers in the economic literature
\cite{Gibrat-1931,Kalecki-1945,Champernowne-1953} use a stochastic
process to describe dynamics of individual wealth or income and to
derive their probability distributions.  One might call this a
one-body approach, because wealth and income fluctuations are
considered independently for each economic agent.  Inspired by
Boltzmann's kinetic theory of collisions in gases, econophysicists
introduced an alternative, two-body approach, where agents perform
pairwise economic transactions and transfer money from one agent to
another.  Actually, this approach was pioneered by the sociologist
John \textcite{Angle-1986,Angle-1992,Angle-1993,Angle-1996,Angle-2002}
already in the 1980s.  However, his work was largely unknown until it
was brought to the attention of econophysicists by the economist
Thomas \textcite{Lux-2005}.  Now, Angle's work is widely cited in
econophysics literature \cite{Angle-2006}.  Meanwhile, the physicists
\textcite{Ispolatov-1998} independently introduced a statistical model
of pairwise money transfer between economic agents, which is
equivalent to the model of Angle.  Soon, three influential papers by
\textcite{Dragulescu-2000,Chakraborti-2000,Bouchaud-2000} appeared and
generated an expanding wave of follow-up publications.  For
pedagogical reasons, we start reviewing this subject with the simplest
version of the pairwise money transfer models presented in
\textcite{Dragulescu-2000}.  This model is the most closely related to
the traditional statistical mechanics, which we briefly review first.
Then we discuss the other models mentioned above, as well as numerous
follow-up papers.

Interestingly, the study of pairwise money transfer and the resulting
statistical distribution of money has virtually no counterpart in
modern economics, so econophysicists initiated a new direction here.
Only the search theory of money \cite{Kiyotaki-1993} is somewhat
related to it.  This theory was an inspiration for the early
econophysics paper by \textcite{Bak-1999} studying dynamics of money.
However, a probability distribution of money among the agents was only
recently obtained within the search-theoretical approach by the
economist Miguel \textcite{Molico-2006}.  His distribution is
qualitatively similar to the distributions found by
\textcite{Angle-1986,Angle-1992,Angle-1993,Angle-1996,Angle-2002,Angle-2006}
and by \textcite{Ispolatov-1998}, but its functional form is unknown,
because it was obtained only numerically.

\subsection{The Boltzmann-Gibbs distribution of energy}
\label{Sec:BGphysics}

The fundamental law of equilibrium statistical mechanics is the
Boltzmann-Gibbs distribution.  It states that the probability
$P(\varepsilon)$ of finding a physical system or subsystem in a state
with the energy $\varepsilon$ is given by the exponential function
\begin{equation}
  P(\varepsilon)=c\,e^{-\varepsilon/T},
\label{Gibbs}
\end{equation}
where $T$ is the temperature, and $c$ is a normalizing constant
\cite{Wannier-book}. Here we set the Boltzmann constant $k_B$ to unity
by choosing the energy units for measuring the physical temperature
$T$.  Then, the expectation value of any physical variable $x$ can be
obtained as
\begin{equation}
  \langle x\rangle=\frac{\sum_k x_ke^{-\varepsilon_k/T}}
  {\sum_k e^{-\varepsilon_k/T}},
\label{expectation}
\end{equation}
where the sum is taken over all states of the system.  Temperature is
equal to the average energy per particle:
$T\sim\langle\varepsilon\rangle$, up to a numerical coefficient of the
order of 1.

Eq.\ (\ref{Gibbs}) can be derived in different ways
\cite{Wannier-book}.  All derivations involve the two main
ingredients: statistical character of the system and conservation of
energy $\varepsilon$.  One of the shortest derivations can be
summarized as follows.  Let us divide the system into two (generally
unequal) parts.  Then, the total energy is the sum of the parts:
$\varepsilon=\varepsilon_1+\varepsilon_2$, whereas the probability is
the product of probabilities:
$P(\varepsilon)=P(\varepsilon_1)\,P(\varepsilon_2)$.  The only
solution of these two equations is the exponential function
(\ref{Gibbs}).

A more sophisticated derivation, proposed by Boltzmann, uses
the concept of entropy.  Let us consider $N$ particles with the total
energy $E$.  Let us divide the energy axis into small intervals (bins)
of width $\Delta\varepsilon$ and count the number of particles $N_k$
having the energies from $\varepsilon_k$ to
$\varepsilon_k+\Delta\varepsilon$.  The ratio $N_k/N=P_k$ gives the
probability for a particle to have the energy $\varepsilon_k$.  Let us
now calculate the multiplicity $W$, which is the number of
permutations of the particles between different energy bins such that
the occupation numbers of the bins do not change.  This quantity is
given by the combinatorial formula in terms of the factorials
\begin{equation}
  W=\frac{N!}{N_1!\,N_2!\,N_3!\,\ldots}.
\label{multiplicity}
\end{equation}
The logarithm of multiplicity is called the entropy $S=\ln W$.  In the
limit of large numbers, the entropy per particle can be written in the
following form using the Stirling approximation for the factorials
\begin{equation}
  \frac{S}{N}=-\sum_k \frac{N_k}{N}\ln\left(\frac{N_k}{N}\right)
  =-\sum_k P_k\ln P_k.
\label{entropy}
\end{equation}
Now we would like to find what distribution of particles among
different energy states has the highest entropy, i.e.,\ the highest
multiplicity, provided the total energy of the system,
$E=\sum_kN_k\varepsilon_k$, has a fixed value.  Solution of this
problem can be easily obtained using the method of Lagrange
multipliers \cite{Wannier-book}, and the answer is given by the
exponential distribution (\ref{Gibbs}).

The same result can be also derived from the ergodic theory, which
says that the many-body system occupies all possible states of a given
total energy with equal probabilities.  Then it is straightforward to
show \cite{Lopez-Ruiz-2008} that the probability distribution of the
energy of an individual particle is given by Eq.\ (\ref{Gibbs}).

\subsection{Conservation of money}
\label{Sec:conservation}

The derivations outlined in Sec.\ \ref{Sec:BGphysics} are very general
and only use the statistical character of the system and the
conservation of energy.  So, one may expect that the exponential
Boltzmann-Gibbs distribution (\ref{Gibbs}) would apply to other
statistical systems with a conserved quantity.

The economy is a big statistical system with millions of participating
agents, so it is a promising target for applications of statistical
mechanics.  Is there a conserved quantity in the economy?
\textcite{Dragulescu-2000} argued that such a conserved quantity is
money $m$.  Indeed, the ordinary economic agents can only receive
money from and give money to other agents.  They are not permitted to
``manufacture'' money, e.g.,\ to print dollar bills.  Let us consider
an economic transaction between agents $i$ and $j$.  When the agent
$i$ pays money $\Delta m$ to the agent $j$ for some goods or services,
the money balances of the agents change as follows
\begin{eqnarray}
  && m_i\;\rightarrow\; m_i'=m_i-\Delta m,
\nonumber \\
  && m_j\;\rightarrow\; m_j'=m_j+\Delta m.
\label{transfer}
\end{eqnarray}
The total amount of money of the two agents before and after
transaction remains the same
\begin{equation}
  m_i+m_j=m_i'+m_j',
\label{conservation}
\end{equation}
i.e.,\ there is a local conservation law for money.  The rule
(\ref{transfer}) for the transfer of money is analogous to the
transfer of energy from one molecule to another in molecular
collisions in a gas, and Eq.\ (\ref{conservation}) is analogous to
conservation of energy in such collisions.  Conservative models of
this kind are also studied in some economic literature
\cite{Kiyotaki-1993,Molico-2006}.

We should emphasize that, in the model of \textcite{Dragulescu-2000}
[as in the economic models of \textcite{Kiyotaki-1993,Molico-2006}],
the transfer of money from one agent to another represents payment for
goods and services in a market economy.  However, the model of
\textcite{Dragulescu-2000} only keeps track of money flow, but does
not keep track of what goods and service are delivered.  One reason
for this is that many goods, e.g.,\ food and other supplies, and most
services, e.g.,\ getting a haircut or going to a movie, are not
tangible and disappear after consumption.  Because they are not
conserved, and also because they are measured in different physical
units, it is not very practical to keep track of them.  In contrast,
money is measured in the same unit (within a given country with a
single currency) and is conserved in local transactions
(\ref{conservation}), so it is straightforward to keep track of money
flow.  It is also important to realize that an increase in material
production does not produce an automatic increase in money supply.
The agents can grow apples on trees, but cannot grow money on trees.
Only a central bank has the monopoly of changing the monetary base
$M_b$ \cite{McConnell-book}.  (Debt and credit issues are discussed
separately in Sec.\ \ref{Sec:debt}.)

Unlike, ordinary economic agents, a central bank or a central
government can inject money into the economy, thus changing the total
amount of money in the system.  This process is analogous to an influx
of energy into a system from external sources, e.g.,\ the Earth
receives energy from the Sun.  Dealing with these situations,
physicists start with an idealization of a closed system in thermal
equilibrium and then generalize to an open system subject to an energy
flux.  As long as the rate of money influx from central sources is
slow compared with relaxation processes in the economy and does not
cause hyperinflation, the system is in quasi-stationary statistical
equilibrium with slowly changing parameters.  This situation is
analogous to heating a kettle on a gas stove slowly, where the kettle
has a well-defined, but slowly increasing, temperature at any moment of
time.  A flux of money may be also produced by international transfers
across the boundaries of a country.  This process involves complicated
issues of multiple currencies in the world and their exchange rates
\cite{McCauley-2008}.  Here we use an idealization of a closed economy
for a single country with a single currency.  Such an idealization is
common in economic literature.  For example, in the two-volume
\textit{Handbook of Monetary Economics} \cite{MonetaryEconomics-book},
only the last chapter out of 23 chapters deals with an open economy.

Another potential problem with conservation of money is debt.  This
issue will be discussed in Sec.\ \ref{Sec:debt}.  As a
starting point, \textcite{Dragulescu-2000} considered simple models,
where debt is not permitted, which is also a common idealization in
some economic literature \cite{Kiyotaki-1993,Molico-2006}.  This means
that money balances of the agents cannot go below zero: $m_i\geq0$ for
all $i$.  Transaction (\ref{transfer}) takes place only when an agent
has enough money to pay the price: $m_i\geq\Delta m$, otherwise the
transaction does not take place.  If an agent spends all money, the
balance drops to zero $m_i=0$, so the agent cannot buy any goods from
other agents.  However, this agent can still receive money from other
agents for delivering goods or services to them.  In real life, money
balance dropping to zero is not at all unusual for people who live
from paycheck to paycheck.

Enforcement of the local conservation law (\ref{conservation}) is the
key feature for successful functioning of money.  If the agents were
permitted to ``manufacture'' money, they would be printing money and
buying all goods for nothing, which would be a disaster.  The physical
medium of money is not essential here, as long as the local
conservation law is enforced.  The days of gold standard are long
gone, so money today is truly the fiat money, declared to be money by
the central bank.  Money may be in the form of paper currency, but
today it is more often represented by digits on computerized bank
accounts.  The local conservation law (\ref{conservation}) is
consistent with the fundamental principles of accounting, whether in
the single-entry or the double-entry form.  More discussion of banks,
debt, and credit will be given in Sec.\ \ref{Sec:debt}.  However, the
macroeconomic monetary policy issues, such as money supply and money
demand \cite{MonetaryEconomics-book}, are outside of the scope of this
paper.  Our goal is to investigate the probability distribution of
money among economic agents.  For this purpose, it is appropriate to
make the simplifying macroeconomic idealizations, as described above,
in order to ensure overall stability of the system and existence of
statistical equilibrium in the model.  The concept of ``equilibrium''
is a very common idealization in economic literature, even though the
real economies might never be in equilibrium.  Here we extend this
concept to a statistical equilibrium, which is characterized by a
stationary probability distribution of money $P(m)$, as opposed to a
mechanical equilibrium, where the ``forces'' of demand and supply
match.

\subsection{The Boltzmann-Gibbs distribution of money}
\label{Sec:BGmoney}

Having recognized the principle of local money conservation,
\textcite{Dragulescu-2000} argued that the stationary distribution of
money $P(m)$ should be given by the exponential Boltzmann-Gibbs
function analogous to Eq.\ (\ref{Gibbs})
\begin{equation}
  P(m)=c\,e^{-m/T_m}.
\label{money}
\end{equation}
Here $c$ is a normalizing constant, and $T_m$ is the ``money
temperature'', which is equal to the average amount of money per
agent: $T=\langle m\rangle=M/N$, where $M$ is the total money, and $N$
is the number of agents.\footnote{Because debt is not permitted in
  this model, we have $M=M_b$, where $M_b$ is the monetary base
  \cite{McConnell-book}.}

\begin{figure}
\includegraphics[width=0.9\linewidth]{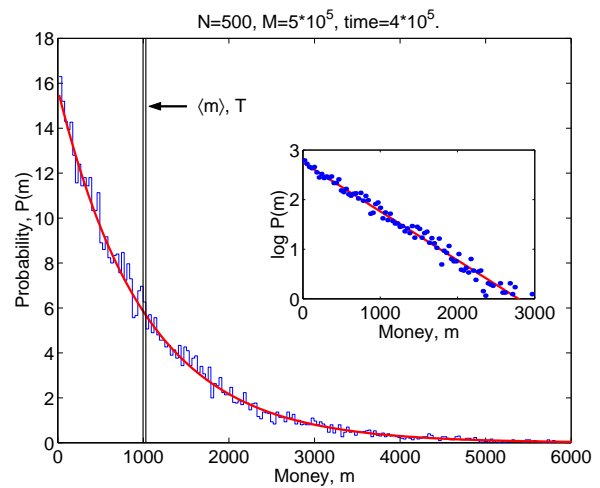}
\caption{\textit{Histogram and points:} Stationary probability
  distribution of money $P(m)$ obtained in agent-based computer
  simulations.  \textit{Solid curves:} Fits to the Boltzmann-Gibbs law
  (\ref{money}).  \textit{Vertical line:} The initial distribution of
  money.  From \textcite{Dragulescu-2000}.}
\label{Fig:money}
\end{figure}

To verify this conjecture, \textcite{Dragulescu-2000} performed
agent-based computer simulations of money transfers between agents.
Initially all agents were given the same amount of money, say, \$1000.
Then, a pair of agents $(i,j)$ was randomly selected, the amount
$\Delta m$ was transferred from one agent to another, and the process
was repeated many times.  Time evolution of the probability
distribution of money $P(m)$ is illustrated in computer animation videos
by \textcite{Chen-2007} and by \textcite{Wright-2007}.  After a
transitory period, money distribution converges to the stationary form
shown in Fig.\ \ref{Fig:money}.  As expected, the distribution is 
well fitted by the exponential function (\ref{money}).

Several different rules for $\Delta m$ were considered by
\textcite{Dragulescu-2000}.  In one model, the transferred amount was
fixed to a constant $\Delta m=\$1$.  Economically, it means that all
agents were selling their products for the same price $\Delta m=\$1$.
Computer animation \cite{Chen-2007} shows that the initial
distribution of money first broadens to a symmetric Gaussian curve,
characteristic for a diffusion process.  Then, the distribution starts
to pile up around the $m=0$ state, which acts as the impenetrable
boundary, because of the imposed condition $m\geq0$.  As a result,
$P(m)$ becomes skewed (asymmetric) and eventually reaches the
stationary exponential shape, as shown in Fig.\ \ref{Fig:money}.  The
boundary at $m=0$ is analogous to the ground-state energy in
statistical physics.  Without this boundary condition, the probability
distribution of money would not reach a stationary state.  Computer
animations \cite{Chen-2007,Wright-2007} also show how the entropy of
money distribution, defined as $S/N=-\sum_k P(m_k)\ln P(m_k)$, grows
from the initial value $S=0$, where all agents have the same money, to
the maximal value at the statistical equilibrium.

While the model with $\Delta m=1$ is very simple and instructive, it
is not realistic, because all prices are taken to be the same.
In another model considered by \textcite{Dragulescu-2000}, $\Delta m$
in each transaction is taken to be a random fraction of the average
amount of money per agent, i.e.,\ $\Delta m=\nu(M/N)$, where $\nu$ is
a uniformly distributed random number between 0 and 1.  The random
distribution of $\Delta m$ is supposed to represent the wide variety
of prices for different products in the real economy.  It reflects the
fact that agents buy and consume many different types of products,
some of them simple and cheap, some sophisticated and expensive.
Moreover, different agents like to consume these products in different
quantities, so there is a variation in the paid amounts $\Delta m$,
even when the unit price of the same product is constant.  Computer
simulation of this model produces exactly the same stationary
distribution (\ref{money}), as in the first model.  Computer animation
for this model is also given by \textcite{Chen-2007}.

The final distribution is universal despite different rules for
$\Delta m$.  To amplify this point further, \textcite{Dragulescu-2000}
also considered a toy model, where $\Delta m$ was taken to be a random
fraction of the average amount of money of the two agents: $\Delta
m=\nu(m_i+m_j)/2$.  This model produced the same stationary
distribution (\ref{money}) as the two other models.

The models of pairwise money transfer are attractive in their
simplicity, but they represent a rather primitive market.  Modern
economy is dominated by big firms, which consist of many agents, so
\textcite{Dragulescu-2000} also studied a model with firms.  One agent
at a time is appointed to become a ``firm''.  The firm borrows capital
$K$ from another agent and returns it with interest $hK$, hires $L$
agents and pays them wages $\omega$, manufactures $Q$ items of a
product, sells them to $Q$ agents at a price $p$, and receives profit
$F=pQ-\omega L-hK$.  All of these agents are randomly selected.  The
parameters of the model are optimized following a procedure from
economics textbooks \cite{McConnell-book}.  The aggregate
demand-supply curve for the product is taken in the form
$p(Q)=v/Q^\eta$, where $Q$ is the quantity consumers would buy at the
price $p$, and $\eta$ and $v$ are some parameters.  The production
function of the firm has the traditional Cobb-Douglas form:
$Q(L,K)=L^\chi K^{1-\chi}$, where $\chi$ is a parameter.  Then the
profit of the firm $F$ is maximized with respect to $K$ and $L$.  The
net result of the firm activity is a many-body transfer of money,
which still satisfies the conservation law.  Computer simulation of
this model generates the same exponential distribution (\ref{money}),
independently of the model parameters.  The reasons for the
universality of the Boltzmann-Gibbs distribution and its limitations
are discussed in Sec.\ \ref{Sec:+x}.

Well after the paper by \textcite{Dragulescu-2000} appeared, the
Italian econophysicists \textcite{Patriarca-2005} found that similar
ideas had been published earlier in obscure Italian journals by
Eleonora \textcite{Bennati-1988,Bennati-1993}.  It was proposed to
call these models the Bennati-Dr\u{a}gulescu-Yakovenko game
\cite{Scalas-2006,Garibaldi-2007}.  The Boltzmann distribution was
independently applied to social sciences by the physicist J\"urgen
\textcite{Mimkes-2000,Mimkes-2005} using the Lagrange principle of
maximization with constraints.  The exponential distribution of money
was also found by the economist Martin \textcite{Shubik-1999} using a
Markov chain approach to strategic market games.  A long time ago,
Benoit \textcite[p 83]{Mandelbrot-1960} observed:
\begin{quote}
  ``There is a great temptation to consider the exchanges of money
  which occur in economic interaction as analogous to the exchanges of
  energy which occur in physical shocks between gas molecules.''
\end{quote}
He realized that this process should result in the exponential
distribution, by analogy with the barometric distribution of density
in the atmosphere.  However, he discarded this idea, because it does
not produce the Pareto power law, and proceeded to study the stable
L\'evy distributions.  Ironically, the actual economic data, discussed
in Secs.\ \ref{Sec:w-empirical} and \ref{Sec:r-data}, do show the
exponential distribution for the majority of the population.
Moreover, the data have a finite variance, so the stable L\'evy
distributions are not applicable because of their infinite variance.

\subsection{Models with debt}
\label{Sec:debt}

Now let us discuss how the results change when debt is
permitted.\footnote{The ideas presented here are quite similar to
  those by \textcite{Soddy-book}.}  From the standpoint of individual
economic agents, debt may be considered as negative money.  When an
agent borrows money from a bank (considered here as a big reservoir of
money),\footnote{Here we treat the bank as being outside of the system
  consisting of ordinary agents, because we are interested in money
  distribution among these agents.  The debt of agents is an asset for
  the bank, and deposits of cash into the bank are liabilities of the
  bank \cite{McConnell-book}.  We do not go into these details in
  order to keep our presentation simple.  For more discussion, see
  \textcite{Keen-2008}.} the cash balance of the agent (positive
money) increases, but the agent also acquires a debt obligation
(negative money), so the total balance (net worth) of the agent
remains the same.  Thus, the act of borrowing money still satisfies a
generalized conservation law of the total money (net worth), which is
now defined as the algebraic sum of positive (cash $M$) and negative
(debt $D$) contributions: $M-D=M_b$.  After spending some cash in
binary transactions (\ref{transfer}), the agent still has the debt
obligation (negative money), so the total money balance $m_i$ of the
agent (net worth) becomes negative.  We see that the boundary
condition $m_i\geq0$, discussed in Sec.\ \ref{Sec:conservation}, does
not apply when debt is permitted, so $m=0$ is not the ground state any
more.  The consequence of permitting debt is not a violation of the
conservation law (which is still preserved in the generalized form for
net worth), but a modification of the boundary condition by permitting
agents to have negative balances $m_i<0$ of net worth.  A more
detailed discussion of positive and negative money and the
book-keeping accounting from the econophysics point of view was
presented by the physicist Dieter \textcite{Braun-2001} and
\textcite{Fischer-2003a,Fischer-2003b}.

Now we can repeat the simulation described in Sec.\ \ref{Sec:BGmoney}
without the boundary condition $m\geq0$ by allowing agents to go into
debt.  When an agent needs to buy a product at a price $\Delta m$
exceeding his money balance $m_i$, the agent is now permitted to
borrow the difference from a bank and, thus, to buy the product.  As a
result of this transaction, the new balance of the agent becomes
negative: $m_i'=m_i-\Delta m<0$.  Notice that the local conservation
law (\ref{transfer}) and (\ref{conservation}) is still satisfied, but
it involves negative values of $m$.  If the simulation is continued
further without any restrictions on the debt of the agents, the
probability distribution of money $P(m)$ never stabilizes, and the
system never reaches a stationary state.  As time goes on, $P(m)$
keeps spreading in a Gaussian manner unlimitedly toward $m=+\infty$
and $m=-\infty$.  Because of the generalized conservation law
discussed above, the first moment $\langle m\rangle=M_b/N$ of the
algebraically defined money $m$ remains constant.  It means that some
agents become richer with positive balances $m>0$ at the expense of
other agents going further into debt with negative balances $m<0$, so
that $M=M_b+D$.

Common sense, as well as the experience with the current financial
crisis, tells us that an economic system cannot be stable if unlimited
debt is permitted.\footnote{In qualitatively agreement with the 
conclusions by \textcite{McCauley-2008}.}  In this case, agents can
buy any goods without producing anything in exchange by simply going
into unlimited debt.  Arguably, the current financial crisis was
caused by the enormous debt accumulation in the system, triggered by
subprime mortgages and financial derivatives based on them.  A widely
expressed opinion is that the current crisis is not the problem of
liquidity, i.e.,\ a temporary difficulty in cash flow, but the problem
of insolvency, i.e.,\ the inherent inability of many participants pay
back their debts.

\begin{figure}
\includegraphics[width=0.9\linewidth]{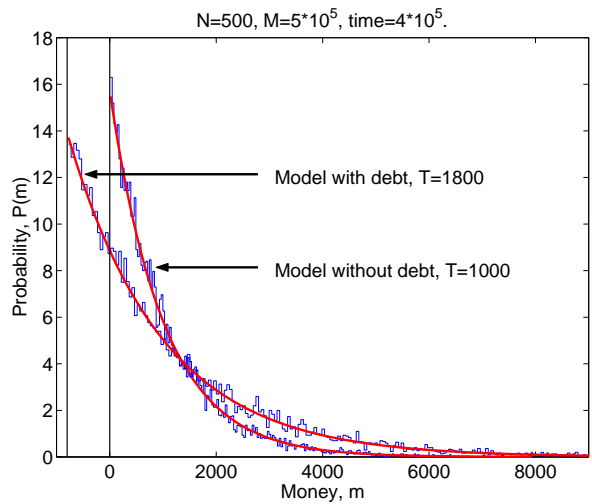}
\caption{\textit{Histograms:} Stationary distributions of money with
  and without debt.  The debt is limited to $m_d=800$. \textit{Solid
  curves:} Fits to the Boltzmann-Gibbs laws with the ``money
  temperatures'' $T_m=1800$ and $T_m=1000$.  From
  \textcite{Dragulescu-2000}.}
\label{Fig:debt}
\end{figure}

Detailed discussion of the current economic situation is not a subject
of this paper.  Going back to the idealized model of money transfers,
one would need to impose some sort of modified boundary conditions in
order to prevent unlimited growth of debt and to ensure overall
stability of the system.  \textcite{Dragulescu-2000} considered a
simple model where the maximal debt of each agent is limited to a
certain amount $m_d$.  This means that the boundary condition
$m_i\geq0$ is now replaced by the condition $m_i\geq-m_d$ for all
agents $i$.  Setting interest rates on borrowed money to be zero for
simplicity, \textcite{Dragulescu-2000} performed computer simulations
of the models described in Sec.\ \ref{Sec:BGmoney} with the new
boundary condition.  The results are shown in Fig.\ \ref{Fig:debt}.
Not surprisingly, the stationary money distribution again has the
exponential shape, but now with the new boundary condition at $m=-m_d$
and the higher money temperature $T_d=m_d+M_b/N$.  By allowing agents
to go into debt up to $m_d$, we effectively increase the amount of
money available to each agent by $m_d$.  So, the money temperature,
which is equal to the average amount of effectively available money
per agent, increases correspondingly.

\begin{figure}
\includegraphics[width=\linewidth]{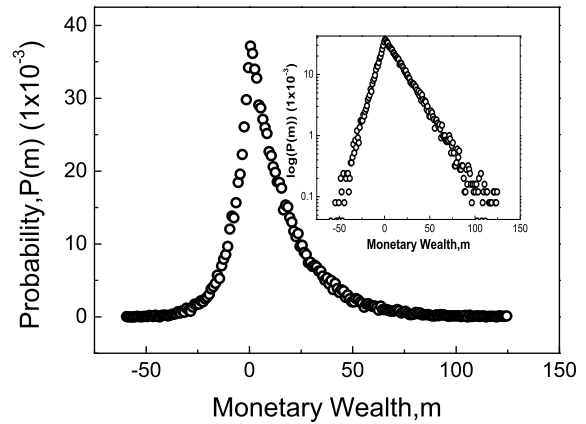}
\caption{The stationary distribution of money for the required reserve
  ratio $R=0.8$.  The distribution is exponential for positive and
  negative money with different ``temperatures'' $T_+$ and $T_-$, as
  illustrated by the inset on log-linear scale.  From
  \textcite{Xi-Ding-Wang-2005}.}
\label{Fig:reserve}
\end{figure}

\textcite{Xi-Ding-Wang-2005} considered another, more realistic
boundary condition, where a constraint is imposed not on the
individual debt of each agent, but on the total debt of all agents in
the system.  This is accomplished via the required reserve ratio $R$,
which is briefly explained below \cite{McConnell-book}.  Banks are
required by law to set aside a fraction $R$ of the money deposited
into bank accounts, whereas the remaining fraction $1-R$ can be loaned
further.  If the initial amount of money in the system (the money
base) is $M_b$, then, with repeated loans and borrowing, the total
amount of positive money available to the agents increases to
$M=M_b/R$, where the factor $1/R$ is called the money multiplier
\cite{McConnell-book}.  This is how ``banks create money''.  Where
does this extra money come from?  It comes from the increase in the
total debt in the system.  The maximal total debt is given by
$D=M_b/R-M_b$ and is limited by the factor $R$.  When the debt is
maximal, the total amounts of positive, $M_b/R$, and negative,
$M_b(1-R)/R$, money circulate among the agents in the system, so there
are two constraints in the model considered by
\textcite{Xi-Ding-Wang-2005}.  Thus, we expect to see the exponential
distributions of positive and negative money characterized by two
different temperatures: $T_+=M_b/RN$ and $T_-=M_b(1-R)/RN$.  This is
exactly what was found in computer simulations by
\textcite{Xi-Ding-Wang-2005}, as shown in Fig.\ \ref{Fig:reserve}.
Similar two-sided distributions were also found by
\textcite{Fischer-2003a}.

However, in reality, the reserve requirement is not effective in
stabilizing total debt in the system, because it applies only to
deposits from general public, but not from corporations
\cite{O'Brian-2007}.\footnote{Australia does not have reserve
  requirements, but China actively uses reserve requirements as a tool
  of monetary policy.}  Moreover, there are alternative instruments of
debt, including derivatives and various unregulated ``financial
innovations''.  As a result, the total debt is not limited in practice
and sometimes can reach catastrophic proportions.  Here we briefly
discuss several models with non-stationary debt.  Thus far, we did not
consider the interest rates.  \textcite{Dragulescu-2000} studied a
simple model with different interest rates for deposits into and loans
from a bank.  Computer simulations found that money distribution among
the agents is still exponential, but the money temperature slowly
changes in time.  Depending on the choice of parameters, the total
amount of money in circulation either increases or decreases in time.
A more sophisticated macroeconomic model was studied by the economist
Steve \textcite{Keen-1995,Keen-2000}.  He found that one of the
regimes is the debt-induced breakdown, where all economic activity
stops under the burden of heavy debt and cannot be restarted without a
``debt moratorium''.  The interest rates were fixed in these models
and not adjusted self-consistently.  \textcite{Cockshott-2008}
proposed a mechanism, where the interest rates are set to cover
probabilistic withdrawals of deposits from a bank.  In an agent-based
simulation of the model, \textcite{Cockshott-2008} found that money
supply first increases up to a certain limit, and then the economy
experiences a spectacular crash under the weight of accumulated debt.
Further studies along these lines would be very interesting.  In the
rest of the paper, we review various models without debt proposed in
literature.

\subsection{Proportional money transfers and saving propensity}
\label{Sec:saving}

\begin{figure}
\includegraphics[width=0.9\linewidth]{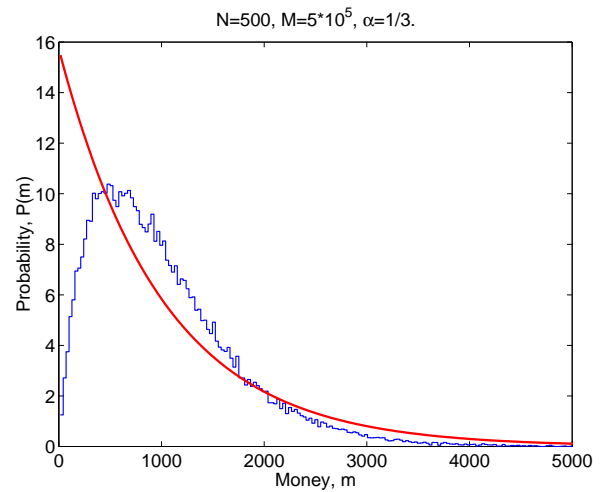}
\caption{\textit{Histogram:} Stationary probability distribution of
  money in the multiplicative random exchange model
  (\ref{proportional}) for $\gamma=1/3$.  \textit{Solid curve:} The
  exponential Boltzmann-Gibbs law.  From
  \textcite{Dragulescu-2000}.}
\label{Fig:Redner}
\end{figure}

In the models of money transfer discussed in Sec.\ \ref{Sec:BGmoney},
the transferred amount $\Delta m$ is typically independent of the
money balances of the agents involved.  A different model was
introduced in physics literature earlier by \textcite{Ispolatov-1998}
and called the multiplicative asset exchange model.  This model also
satisfies the conservation law, but the transferred amount of money is
a fixed fraction $\gamma$ of the payer's money in
Eq.\ (\ref{transfer}):
\begin{equation}
  \Delta m=\gamma m_i.
\label{proportional}
\end{equation}
The stationary distribution of money in this model, compared in
Fig.\ \ref{Fig:Redner} with an exponential function, is similar, but
not exactly equal, to the Gamma distribution:
\begin{equation}
  P(m)=c\,m^\beta\,e^{-m/T}.
\label{Gamma}
\end{equation}
Eq.\ (\ref{Gamma}) differs from Eq.\ (\ref{money}) by the power-law
prefactor $m^\beta$.  From the Boltzmann kinetic equation (discussed 
in Sec.\ \ref{Sec:+x}), \textcite{Ispolatov-1998}
derived a formula relating the parameters $\gamma$ and $\beta$ in
Eqs.\ (\ref{proportional}) and (\ref{Gamma}):
\begin{equation}
  \beta=-1-\ln2/\ln(1-\gamma).
\label{beta}
\end{equation}
When payers spend a relatively small fraction of their money
$\gamma<1/2$, Eq.\ (\ref{beta}) gives $\beta>0$.  In this case, the
population with low money balances is reduced, and $P(0)=0$, as shown
in Fig.\ \ref{Fig:Redner}.

The economist Thomas \textcite{Lux-2005} brought to the attention of
physicists that essentially the same model, called the inequality
process, had been introduced and studied much earlier by the
sociologist John
\textcite{Angle-1986,Angle-1992,Angle-1993,Angle-1996,Angle-2002}, see
also the review by \textcite{Angle-2006} for additional references.
While \textcite{Ispolatov-1998} did not give much justification for
the proportionality law (\ref{proportional}), \textcite{Angle-1986}
connected this rule with the surplus theory of social stratification
\cite{Engels-book}, which argues that inequality in human society
develops when people can produce more than necessary for minimal
subsistence.  This additional wealth (surplus) can be transferred from
original producers to other people, thus generating inequality.  In
the first paper by \textcite{Angle-1986}, the parameter $\gamma$ was
randomly distributed, and another parameter $\delta$ gave a higher
probability of winning to the agent with the higher money balance in
Eq.\ (\ref{transfer}).  However, in the following papers, he
simplified the model to a fixed $\gamma$ (denoted as $\omega$ by
Angle) and equal probabilities of winning for higher- and
lower-balance agents, which makes it completely equivalent to the
model of \textcite{Ispolatov-1998}.  \textcite{Angle-2002,Angle-2006}
also considered a model where groups of agents have different values
of $\gamma$, simulating the effect of education and other ``human
capital''.  All of these models generate a Gamma-like distribution,
well approximated by Eq.\ (\ref{Gamma}).

Another model with an element of proportionality was proposed by
\textcite{Chakraborti-2000}.\footnote{This paper originally appeared
  as a follow-up e-print cond-mat/0004256 on the e-print
  cond-mat/0001432 by \textcite{Dragulescu-2000}.}  In this model, the
agents set aside (save) some fraction of their money $\lambda m_i$,
whereas the rest of their money balance $(1-\lambda)m_i$ becomes
available for random exchanges.  Thus, the rule of exchange
(\ref{transfer}) becomes
\begin{eqnarray}
  && m_i'=\lambda m_i + \xi(1-\lambda)(m_i+m_j),
\nonumber \\
  && m_j'=\lambda m_j + (1-\xi)(1-\lambda)(m_i+m_j).
\label{saving}
\end{eqnarray}
Here the coefficient $\lambda$ is called the saving propensity, and
the random variable $\xi$ is uniformly distributed between 0 and 1.
It was pointed out by \textcite{Angle-2006} that, by the change of
notation $\lambda\to(1-\gamma)$, Eq.\ (\ref{saving}) can be
transformed to the same form as Eq.\ (\ref{proportional}), if the
random variable $\xi$ takes only discrete values 0 and 1.  Computer
simulations by \textcite{Chakraborti-2000} of the model (\ref{saving})
found a stationary distribution close to the Gamma distribution
(\ref{Gamma}).  It was shown that the parameter $\beta$ is related to
the saving propensity $\lambda$ by the formula
$\beta=3\lambda/(1-\lambda)$
\cite{Patriarca-2004a,Patriarca-2004b,Patriarca-2005,Repetowicz-2005}.
For $\lambda\neq0$, agents always keep some money, so their balances
never drop to zero, and $P(0)=0$, whereas for $\lambda=0$ the
distribution becomes exponential.

In the subsequent papers by the Kolkata school
\cite{Chakrabarti-history} and related papers, the case of random
saving propensity was studied.  In these models, the agents are
assigned random parameters $\lambda$ drawn from a uniform distribution
between 0 and 1 \cite{Chatterjee-2004}.  It was found that this model
produces a power-law tail $P(m)\propto1/m^2$ at high $m$.  The reasons
for stability of this law were understood using the Boltzmann kinetic
equation \cite{Das-2005,Chatterjee-2005,Repetowicz-2005}, but most
elegantly in the mean-field theory
\cite{Mohanty-2006,Bhattacharyya-2007,Chatterjee-2007}.  The fat tail
originates from the agents whose saving propensity is close to 1, who
hoard money and do not give it back
\cite{Patriarca-2005,Patriarca-2006}.  A more rigorous mathematical
treatment of the problem was given by
\textcite{During-2008,Matthes-2008,During-2007}.  An interesting
matrix formulation of the problem was presented by
\textcite{Gupta-2006}.  Relaxation rate in the money transfer models
was studied by \textcite{During-2008,Patriarca-2007,Gupta-2008}.
\textcite{Dragulescu-2000} considered a model with taxation, which
also has an element of proportionality.  The Gamma distribution was
also studied for conservative models within a simple Boltzmann
approach by \textcite{Ferrero-2004} and, using more complicated rules
of exchange motivated by political economy, by
\textcite{Scafetta-2004a,Scafetta-2004b}.  Independently, the
economist Miguel \textcite{Molico-2006} studied conservative exchange
models where agents bargain over prices in their transactions.  He
found stationary Gamma-like distributions of money in numerical
simulations of these models.

\subsection{Additive versus multiplicative models}
\label{Sec:+x}

The stationary distribution of money (\ref{Gamma}) for the models of
Sec.\ \ref{Sec:saving} is different from the simple exponential
formula (\ref{money}) found for the models of Sec.\ \ref{Sec:BGmoney}.
The origin of this difference can be understood from the Boltzmann
kinetic equation \cite{Wannier-book,Kinetics-book}.  This equation
describes time evolution of the distribution function $P(m)$ due to
pairwise interactions:
\begin{eqnarray}
  &&\frac{dP(m)}{dt}=\int\!\!\!\!\int\{
    -f_{[m,m']\to[m-\Delta,m'+\Delta]}P(m)P(m')
\label{Boltzmann}  \\
  &&+f_{[m-\Delta,m'+\Delta]\to[m,m']}
  P(m-\Delta)P(m'+\Delta)\}\,dm'\,d\Delta.
\nonumber
\end{eqnarray}
Here $f_{[m,m']\to[m-\Delta,m'+\Delta]}$ is the probability of
transferring money $\Delta$ from an agent with money $m$ to an agent
with money $m'$ per unit time.  This probability, multiplied by the
occupation numbers $P(m)$ and $P(m')$, gives the rate of transitions
from the state $[m,m']$ to the state $[m-\Delta,m'+\Delta]$.  The
first term in Eq.\ (\ref{Boltzmann}) gives the depopulation rate of
the state $m$.  The second term in Eq.\ (\ref{Boltzmann}) describes
the reversed process, where the occupation number $P(m)$ increases.
When the two terms are equal, the direct and reversed transitions
cancel each other statistically, and the probability distribution is
stationary: $dP(m)/dt=0$.  This is the principle of detailed balance.

In physics, the fundamental microscopic equations of motion obey the
time-reversal symmetry.  This means that the probabilities of the
direct and reversed processes are exactly equal:
\begin{eqnarray}
  f_{[m,m']\to[m-\Delta,m'+\Delta]}=f_{[m-\Delta,m'+\Delta]\to[m,m']}.
\label{reversal}
\end{eqnarray}
When Eq.\ (\ref{reversal}) is satisfied, the detailed balance
condition for Eq.\ (\ref{Boltzmann}) reduces to the equation
$P(m)P(m')=P(m-\Delta)P(m'+\Delta)$, because the factors $f$ cancels
out.  The only solution of this equation is the exponential function
$P(m)=c\exp(-m/T_m)$, so the Boltzmann-Gibbs distribution is the
stationary solution of the Boltzmann kinetic equation
(\ref{Boltzmann}).  Notice that the transition probabilities
(\ref{reversal}) are determined by the dynamical rules of the model,
but the equilibrium Boltzmann-Gibbs distribution does not depend on
the dynamical rules at all.  This is the origin of the universality of
the Boltzmann-Gibbs distribution.  We see that it is possible to find
the stationary distribution without knowing details of the dynamical
rules (which are rarely known very well), as long as the symmetry
condition (\ref{reversal}) is satisfied.

The models considered in Sec.\ \ref{Sec:BGmoney} have the
time-reversal symmetry.  The model with the fixed money transfer
$\Delta$ has equal probabilities (\ref{reversal}) of transferring
money from an agent with the balance $m$ to an agent with the balance
$m'$ and vice versa.  This is also true when $\Delta$ is random, as
long as the probability distribution of $\Delta$ is independent of $m$
and $m'$.  Thus, the stationary distribution $P(m)$ is always
exponential in these models.

However, there is no fundamental reason to expect the time-reversal
symmetry in economics, where Eq.\ (\ref{reversal}) may be not valid.
In this case, the system may have a non-exponential stationary
distribution or no stationary distribution at all.  In the model
(\ref{proportional}), the time-reversal symmetry is broken.  Indeed,
when an agent $i$ gives a fixed fraction $\gamma$ of his money $m_i$
to an agent with balance $m_j$, their balances become $(1-\gamma)m_i$
and $m_j+\gamma m_i$.  If we try to reverse this process and appoint
the agent $j$ to be the payer and to give the fraction $\gamma$ of her
money, $\gamma(m_j+\gamma m_i)$, to the agent $i$, the system does not
return to the original configuration $[m_i,m_j]$.  As emphasized by
\textcite{Angle-2006}, the payer pays a deterministic fraction of his
money, but the receiver receives a random amount from a random agent,
so their roles are not interchangeable.  Because the proportional rule
typically violates the time-reversal symmetry, the stationary
distribution $P(m)$ in multiplicative models is typically not
exponential.\footnote{However, when $\Delta m$ is a fraction of the
  total money $m_i+m_j$ of the two agents, the model is
  time-reversible and has the exponential distribution, as discussed
  in Sec.\ \ref{Sec:BGmoney}.}  Making the transfer dependent on the
money balance of the payer effectively introduces Maxwell's demon into
the model.  Another view on the time-reversal symmetry in economic
dynamics was presented by \textcite{Ao-2007}.

These examples show that the Boltzmann-Gibbs distribution does not
necessarily hold for any conservative model.  However, it is universal
in a limited sense.  For a broad class of models that have
time-reversal symmetry, the stationary distribution is exponential and
does not depend on details of a model.  Conversely, when the
time-reversal symmetry is broken, the distribution may depend on
details of a model.  The difference between these two classes of
models may be rather subtle.  Deviations from the Boltzmann-Gibbs law
may occur only if the transition rates $f$ in Eq.\ (\ref{reversal})
explicitly depend on the agents' money $m$ or $m'$ in an asymmetric
manner.  \textcite{Dragulescu-2000} performed a computer simulation
where the direction of payment was randomly fixed in advance for every
pair of agents $(i,j)$.  In this case, money flows along directed
links between the agents: $i\!\to\!j\!\to\!k$, and the time-reversal
symmetry is strongly violated.  This model is closer to the real
economy, where one typically receives money from an employer and pays
it to a grocery store.  Nevertheless, the Boltzmann-Gibbs distribution
was still found in this model, because the transition rates $f$ do not
explicitly depend on $m$ and $m'$ and do not violate
Eq.\ (\ref{reversal}).  A more general study of money exchange models
on directed networks was presented by \textcite{Chatterjee-2009}.

In the absence of detailed knowledge of real microscopic dynamics of
economic exchanges, the semiuniversal Boltzmann-Gibbs distribution
(\ref{money}) is a natural starting point.  Moreover, the assumption
of \textcite{Dragulescu-2000} that agents pay the same prices $\Delta
m$ for the same products, independent of their money balances $m$,
seems very appropriate for the modern anonymous economy, especially
for purchases over the Internet.  There is no particular empirical
evidence for the proportional rules (\ref{proportional}) or
(\ref{saving}).  However, the difference between the additive
(\ref{money}) and multiplicative (\ref{Gamma}) distributions may be
not so crucial after all.  From the mathematical point of view, the
difference is in the implementation of the boundary condition at
$m=0$.  In the additive models of Sec.\ \ref{Sec:BGmoney}, there is a
sharp cutoff for $P(m)\neq0$ at $m=0$.  In the multiplicative models
of Sec.\ \ref{Sec:saving}, the balance of an agent never reaches
$m=0$, so $P(m)$ vanishes at $m\to0$ in a power-law manner.  But for
large $m$, $P(m)$ decreases exponentially in both models.

By further modifying the rules of money transfer and introducing more
parameters in the models, one can obtain even more complicated
distributions \cite{Scafetta-2007,Saif-2007}.  However, one can argue
that parsimony is the virtue of a good mathematical model, not the
abundance of additional assumptions and parameters, whose
correspondence to reality is hard to verify.

\section{Statistical Mechanics of Wealth Distribution}
\label{Sec:wealth}

In the econophysics literature on exchange models, the terms ``money''
and ``wealth'' are often used interchangeably.  However, economists
emphasize the difference between these two concepts.  In this section,
we review the models of wealth distribution, as opposed to money
distribution.

\subsection{Models with a conserved commodity}
\label{Sec:w=const}

What is the difference between money and wealth?
\textcite{Dragulescu-2000} argued that wealth $w_i$ is equal to money
$m_i$ plus the other property that an agent $i$ has.  The latter may
include durable material property, such as houses and cars, and
financial instruments, such as stocks, bonds, and options.  Money
(paper cash, bank accounts) is generally liquid and countable.
However, the other property is not immediately liquid and has to be
sold first (converted into money) to be used for other purchases.  In
order to estimate the monetary value of property, one needs to know
its price $p$.  In the simplest model, let us consider just one type
of property, say, stocks $s$.  Then the wealth of an agent $i$ is
given by
\begin{equation}
  w_i=m_i+p\,s_i.
\label{wealth}
\end{equation}
It is assumed that the price $p$ is common for all agents and is
established by some kind of market process, such as an auction, and
may change in time.

It is reasonable to start with a model where both the total money
$M=\sum_im_i$ and the total stock $S=\sum_is_i$ are conserved
\cite{Chakraborti-2001,Chatterjee-2006,Ausloos-2007}.  The agents pay
money to buy stock and sell stock to get money, and so on.  Although
$M$ and $S$ are conserved, the total wealth $W=\sum_iw_i$ is generally
not conserved \cite{Chatterjee-2006}, because of price fluctuation in
Eq.\ (\ref{wealth}).  This is an important difference from the money
transfers models of Sec.\ \ref{Sec:money}.  The wealth $w_i$ of an
agent $i$, not participating in any transactions, may change when
transactions between other agents establish a new price $p$.
Moreover, the wealth $w_i$ of an agent $i$ does not change after a
transaction with an agent $j$.  Indeed, in exchange for paying money
$\Delta m$, the agent $i$ receives the stock $\Delta s=\Delta m/p$, so
her total wealth (\ref{wealth}) remains the same.  Theoretically, the
agent can instantaneously sell the stock back at the same price and
recover the money paid.  If the price $p$ never changes, then the
wealth $w_i$ of each agent remains constant, despite transfers of
money and stock between agents.

We see that redistribution of wealth in this model is directly related
to price fluctuations.  A mathematical model of this process was
studied by \textcite{Silver-2002}.  In this model, the agents randomly
change preferences for the fraction of their wealth invested in
stocks.  As a result, some agents offer stock for sale and some want
to buy it.  The price $p$ is determined from the market-clearing
auction matching supply and demand.  \textcite{Silver-2002}
demonstrated in computer simulations and proved analytically using the
theory of Markov processes that the stationary distribution $P(w)$ of
wealth $w$ in this model is given by the Gamma distribution, as in
Eq.\ (\ref{Gamma}).  Various modifications of this model considered by
\textcite{Lux-2005}, such as introducing monopolistic coalitions, do
not change this result significantly, which shows robustness of the
Gamma distribution.  For models with a conserved commodity,
\textcite{Chatterjee-2006} found the Gamma distribution for a fixed
saving propensity and a power-law tail for a distributed saving
propensity.

Another model with conserved money and stock was studied by
\textcite{Raberto-2003} for an artificial stock market, where traders
follow different investment strategies: random, momentum, contrarian,
and fundamentalist.  Wealth distribution in the model with random
traders was found have a power-law tail $P(w)\sim1/w^2$ for large $w$.
However, unlike in other simulations, where all agents initially
have equal balances, here the initial money and stock balances of the
agents were randomly populated according to a power law with the same
exponent.  This raises the question whether the observed power-law
distribution of wealth is an artifact of the initial conditions,
because equilibration of the upper tail may take a very long
simulation time.

\subsection{Models with stochastic growth of wealth}
\label{Sec:w-not-const}

Although the total wealth $W$ is not exactly conserved in the models
considered in Sec.\ \ref{Sec:w=const}, nevertheless $W$ remains
constant on average, because the total money $M$ and stock $S$ are
conserved.  A different model for wealth distribution was proposed by
\textcite{Bouchaud-2000}.  In this model, time evolution of the wealth
$w_i$ of an agent $i$ is given by the stochastic differential equation
\begin{equation}
  \frac{dw_i}{dt}=\eta_i(t)\,w_i + \sum_{j(\neq i)} J_{ij}w_j -
  \sum_{j(\neq i)} J_{ji}w_i,
\label{Bouchaud}
\end{equation}
where $\eta_i(t)$ is a Gaussian random variable with the mean
$\langle\eta\rangle$ and the variance $2\sigma^2$.  This variable
represents growth or loss of wealth of an agent due to investment in
stock market.  The last two terms describe transfer of wealth between
different agents, which is taken to be proportional to the wealth of
the payers with the coefficients $J_{ij}$.  So, the model
(\ref{Bouchaud}) is multiplicative and invariant under the scale
transformation $w_i\to Zw_i$.  For simplicity, the exchange fractions
are taken to be the same for all agents: $J_{ij}=J/N$ for all $i\neq
j$, where $N$ is the total number of agents.  In this case, the last
two terms in Eq.\ (\ref{Bouchaud}) can be written as $J(\langle
w\rangle - w_i)$, where $\langle w\rangle=\sum_iw_i/N$ is the average
wealth per agent.  This case represents a ``mean-field'' model, where
all agents feel the same environment.  It can be easily shown that the
average wealth increases in time as $\langle w\rangle_t=\langle
w\rangle_0e^{(\langle\eta\rangle+\sigma^2)t}$.  Then, it makes more
sense to consider the relative wealth $\tilde w_i=w_i/\langle
w\rangle_t$.  Eq.\ (\ref{Bouchaud}) for this variable becomes
\begin{equation}
  \frac{d\tilde
  w_i}{dt}=(\eta_i(t)-\langle\eta\rangle-\sigma^2)\,\tilde w_i
  +J(1-\tilde w_i).
\label{relative}
\end{equation}
The probability distribution $P(\tilde w,t)$ for the stochastic
differential equation (\ref{relative}) is governed by the
Fokker-Planck equation
\begin{equation}
  \frac{\partial P}{\partial t}
  =\frac{\partial[J(\tilde w-1)+\sigma^2\tilde w] P}{\partial \tilde w}
  +\sigma^2\frac{\partial}{\partial\tilde w}
  \left(\tilde w\frac{\partial(\tilde wP)}{\partial\tilde w}\right).
\label{FP-Bouchaud}
\end{equation}
The stationary solution ($\partial P/\partial t=0$) of this equation
is given by the following formula
\begin{equation}
  P(\tilde w)=c\,\frac{e^{-J/\sigma^2\tilde w}}
  {\tilde w^{2+J/\sigma^2}}.
\label{P-Bouchaud}
\end{equation}
The distribution (\ref{P-Bouchaud}) is quite different from the
Boltzmann-Gibbs (\ref{money}) and Gamma (\ref{Gamma}) distributions.
Eq.\ (\ref{P-Bouchaud}) has a power-law tail at large $\tilde w$ and a
sharp cutoff at small $\tilde w$.  Eq.\ (\ref{Bouchaud}) is a version
of the generalized Lotka-Volterra model, and the stationary
distribution (\ref{P-Bouchaud}) was also obtained by
\textcite{Solomon-2001,Solomon-2002}.  The model was generalized to
include negative wealth by \textcite{Huang-2004}.

\textcite{Bouchaud-2000} used the mean-field approach.  A similar
result was found for a model with pairwise interaction between agents
by \textcite{Slanina-2004}.  In his model, wealth is transferred
between the agents following the proportional rule
(\ref{proportional}), but, in addition, the wealth of the agents
increases by the factor $1+\zeta$ in each transaction.  This factor is
supposed to reflect creation of wealth in economic
interactions. Because the total wealth in the system increases, it
makes sense to consider the distribution of relative wealth $P(\tilde
w)$.  In the limit of continuous trading, \textcite{Slanina-2004}
found the same stationary distribution (\ref{P-Bouchaud}).  This
result was reproduced using a mathematically more involved treatment
of this model by \textcite{Cordier-2005,Pareschi-2006}.  Numerical
simulations of the models with stochastic noise $\eta$ by
\textcite{Scafetta-2004a,Scafetta-2004b} also found a power law tail
for large $w$.  Equivalence between the models with pairwise exchange
and exchange with a reservoir was discussed by \textcite{Basu-2008}.

We now contrast the models discussed in Secs.\ \ref{Sec:w=const} and
\ref{Sec:w-not-const}.  In the former case, where money and commodity
are conserved, and wealth does not grow, the distribution of wealth is
given by the Gamma distribution with the exponential tail for large
$w$.  In the latter models, wealth grows in time exponentially, and
the distribution of relative wealth has a power-law tail for large
$\tilde w$.  These results suggest that the presence of a power-law
tail is a nonequilibrium effect that requires constant growth or
inflation of the economy, but disappears for a closed system with
conservation laws.

The discussed models were reviewed by
\textcite{Richmond-2006a,Richmond-2006b,Chatterjee-2007,Yakovenko-2009}
and in the popular article by \textcite{Hayes-2002}.  Because of lack
of space, we omit discussion of models with wealth condensation
\cite{Bouchaud-2000,Ispolatov-1998,Burda-2002,Pianegonda-2003,Braun-2006},
where a few agents accumulate a finite fraction of the total wealth,
and studies of wealth distribution on complex networks
\cite{Coelho-2005,Iglesias-2003,DiMatteo-2004,Hu-2006,Hu-2007}.  So
far, we discussed the models with long-range interaction, where any
agent can exchange money and wealth with any other agent.  A local
model, where agents trade only with the nearest neighbors, was studied
by \textcite{Bak-1999}.

\subsection{Empirical data on money and wealth distributions}
\label{Sec:w-empirical}

\begin{figure}
\includegraphics[width=0.9\linewidth]{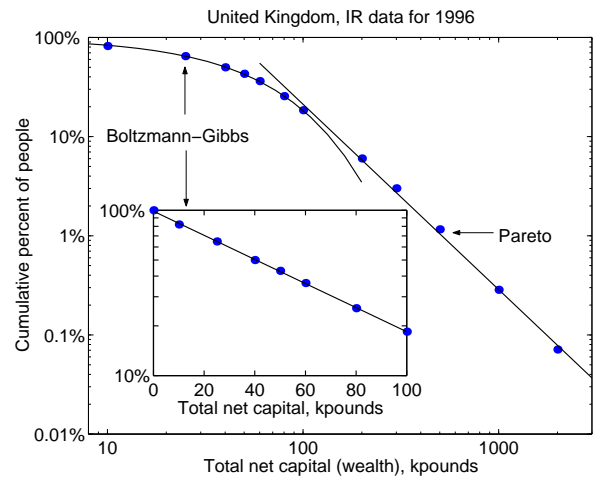}
\caption{Cumulative probability distribution of net wealth in the UK
  shown on log-log (main panel) and log-linear (inset) scales.  Points
  represent the data from the Inland Revenue, and solid lines are fits
  to the exponential (Boltzmann-Gibbs) and power (Pareto) laws.
  From \textcite{Dragulescu-2001b}.}
\label{Fig:UKwealth}
\end{figure}

It would be interesting to compare theoretical results for money
and wealth distributions in various models with empirical data.
Unfortunately, such empirical data are difficult to find.  Unlike
income, which is discussed in Sec.\ \ref{Sec:income}, wealth is not
routinely reported by the majority of individuals to the government.
However, in some countries, when a person dies, all assets must be
reported for the purpose of inheritance tax.  So, in principle, there
exist good statistics of wealth distribution among dead people, which,
of course, is different from the wealth distribution among the living.  
Using an adjustment procedure based on the age, gender, and
other characteristics of the deceased, the UK tax agency, the Inland
Revenue, reconstructed the wealth distribution of the whole population
of the UK \cite{UKwealth}.  Fig.\ \ref{Fig:UKwealth} shows the UK data
for 1996 reproduced from \textcite{Dragulescu-2001b}.  The figure
shows the cumulative probability $C(w)=\int_w^\infty P(w')\,dw'$ as a
function of the personal net wealth $w$, which is composed of assets
(cash, stocks, property, household goods, etc.)\ and liabilities
(mortgages and other debts).  Because statistical data are usually
reported at non-uniform intervals of $w$, it is more practical to plot
the cumulative probability distribution $C(w)$ rather than its
derivative, the probability density $P(w)$.  Fortunately, when $P(w)$
is an exponential or a power-law function, then $C(w)$ is also an
exponential or a power-law function.

The main panel in Fig.\ \ref{Fig:UKwealth} shows a plot of $C(w)$ on
the log-log scale, where a straight line represents a power-law
dependence.  The figure shows that the distribution follows a power
law $C(w)\propto1/w^\alpha$ with the exponent $\alpha=1.9$ for the
wealth greater than about 100~k$\pounds$.  The inset in
Fig.\ \ref{Fig:UKwealth} shows the same data on the log-linear scale,
where a straight line represents an exponential dependence.  We
observe that, below 100~k$\pounds$, the data are well fitted by the
exponential distribution $C(w)\propto\exp(-w/T_w)$ with the effective
``wealth temperature'' $T_w=60$~k$\pounds$ (which corresponds to the
median wealth of 41~k$\pounds$).  So, the distribution of wealth is
characterized by the Pareto power law in the upper tail of the
distribution and the exponential Boltzmann-Gibbs law in the lower part
of the distribution for the great majority (about 90\%) of the
population.  Similar results are found for the distribution of income,
as discussed in Sec.\ \ref{Sec:income}.  One may speculate that wealth
distribution in the lower part is dominated by distribution of money,
because the corresponding people do not have other significant assets
\cite{Levy-Levy-2003}, so the results of Sec.\ \ref{Sec:money} give
the Boltzmann-Gibbs law.  On the other hand, the upper tail of wealth
distribution is dominated by investment assess \cite{Levy-Levy-2003},
where the results of Sec.\ \ref{Sec:w-not-const} give the Pareto law.
The power law was studied by many researchers
\cite{Levy-2003,Levy-Levy-2003,Klass-2007,Sinha-2006} for the
upper-tail data, such as the Forbes list of 400 richest people.  On
the other hand, statistical surveys of the population, such as the
Survey of Consumer Finance \cite{Diaz-Gimenez-1997} and the Panel
Study of Income Dynamics (PSID), give more information about the lower
part of the wealth distribution.  Curiously, \textcite{Abul-Magd-2002}
found that the wealth distribution in the ancient Egypt was consistent
with Eq.\ (\ref{P-Bouchaud}).  \textcite{Hegyi-2007} found a power-law
tail for the wealth distribution of aristocratic families in medieval
Hungary.

For direct comparison with the results of Sec.\ \ref{Sec:money}, it
would be interesting to find data on the distribution of money,
as opposed to the distribution of wealth.  Making a reasonable
assumption that most people keep most of their money in banks, one can
approximate the distribution of money by the distribution of balances
on bank accounts.  (Balances on all types of bank accounts, such as
checking, saving, and money manager, associated with the same person
should be added up.)  Despite imperfections (people may have accounts
in different banks or not keep all their money in banks), the
distribution of balances on bank accounts would give valuable
information about the distribution of money.  The data for a large
enough bank would be representative of the distribution in the whole
economy.  Unfortunately, it has not been possible to obtain such data
thus far, even though it would be completely anonymous and not
compromise privacy of bank clients.

The data on the distribution of bank accounts balances would be
useful, e.g.,\ to the Federal Deposits Insurance Company (FDIC) of the
USA.  This government agency insures bank deposits of customers up to
a certain maximal balance.  In order to estimate its exposure and the
change in exposure due to a possible increase in the limit, FDIC would
need to know the probability distribution of balances on bank
accounts.  It is quite possible that FDIC may already have such data.

\begin{figure}
\includegraphics[width=0.9\linewidth]{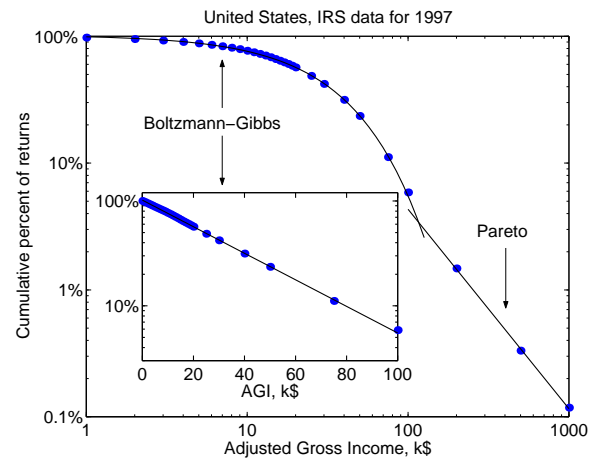}
\caption{Cumulative probability distribution of tax returns for USA in
  1997 shown on log-log (main panel) and log-linear (inset) scales.
  Points represent the Internal Revenue Service data, and solid lines
  are fits to the exponential and power-law functions.  From 
  \textcite{Dragulescu-2003}.}
\label{Fig:income1997}
\end{figure}

Measuring the probability distribution of money would be also very
useful for determining how much people can, in principle, spend on
purchases (without going into debt).  This is different from the
distribution of wealth, where the property component, such as a house,
a car, or retirement investment, is effectively locked up and, in most
cases, is not easily available for consumer spending.  Thus, although
wealth distribution may reflect the distribution of economic power,
the distribution of money is more relevant for immediate consumption.

\section{Data and Models for Income Distribution}
\label{Sec:income}

In contrast to money and wealth distributions, more empirical
data are available for the distribution of income $r$ from tax
agencies and population surveys.  In this section, we first present
empirical data on income distribution and then discuss theoretical
models.

\subsection{Empirical data on income distribution}
\label{Sec:r-data}

Empirical studies of income distribution have a long history in the
economic literature.\footnote{See, e.g.,\ 
\textcite{Kakwani-book,Champernowne-1998,Atkinson-2000,Atkinson-2007,Piketty-2003}.}
Many articles on this subject appear in the journal \textit{Review of
Income and Wealth}, published on behalf of the International
Association for Research in Income and Wealth.  Following the work by
\textcite{Pareto-book}, much attention was focused on the power-law
upper tail of income distribution and less on the lower part.  In
contrast to more complicated functions discussed in the economic
literature \cite{Kakwani-book,Champernowne-1998,Atkinson-2000},
\textcite{Dragulescu-2001a} demonstrated that the lower part of income
distribution can be well fitted with the simple exponential function
$P(r)=c\exp(-r/T_r)$, which is characterized by just one parameter,
the ``income temperature'' $T_r$.  Then
\textcite{Dragulescu-2001b,Dragulescu-2003} showed that the whole
income distribution can be fitted by an exponential function in the
lower part and a power-law function in the upper part, as shown in
Fig.\ \ref{Fig:income1997}.  The straight line on the log-linear scale
in the inset of Fig.\ \ref{Fig:income1997} demonstrates the
exponential Boltzmann-Gibbs law, and the straight line on the log-log
scale in the main panel illustrates the Pareto power law.  The fact
that income distribution consists of two distinct parts reveals the
two-class structure of the American society
\cite{Yakovenko-2005,Silva-2005}.  Coexistence of the exponential and
power-law distributions is also known in plasma physics and
astrophysics, where they are called the ``thermal'' and
``superthermal'' parts \cite{Hasegawa-1985,Desai-2003,Collier-2004}.
The boundary between the lower and upper classes can be defined as the
intersection point of the exponential and power-law fits in
Fig.\ \ref{Fig:income1997}.  For 1997, the annual income separating
the two classes was about 120~k\$.  About 3\% of the population
belonged to the upper class, and 97\% belonged to the lower class.

\begin{figure}
\includegraphics[angle=-90,width=0.95\linewidth]{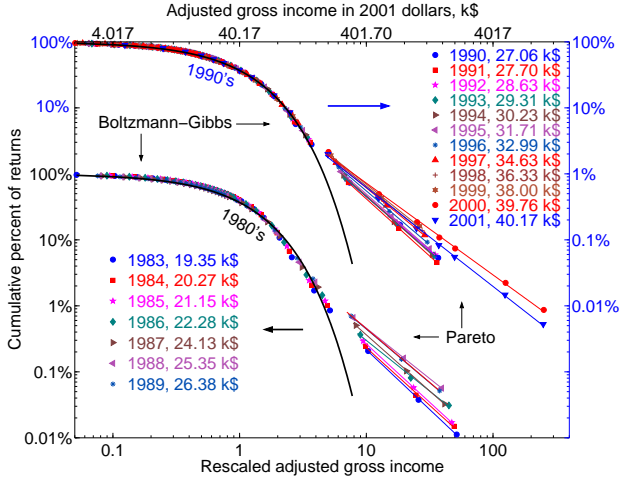}
\caption{Cumulative probability distribution of tax returns plotted on
  log-log scale versus $r/T_r$ (the annual income $r$ normalized by
  the average income $T_r$ in the exponential part of the
  distribution).  The IRS data points are for 1983--2001, and the
  columns of numbers give the values of $T_r$ for the corresponding
  years.  From \textcite{Silva-2005}.}
\label{Fig:income}
\end{figure}

\textcite{Silva-2005} studied time evolution of income distribution in
the USA during 1983--2001 using the data from the Internal Revenue
Service (IRS), the government tax agency.  The structure of income
distribution was found to be qualitatively the same for all years, as
shown in Fig.\ \ref{Fig:income}.  The average income in nominal
dollars has approximately doubled during this time interval.  So, the
horizontal axis in Fig.\ \ref{Fig:income} shows the normalized income
$r/T_r$, where the ``income temperature'' $T_r$ was obtained by
fitting of the exponential part of the distribution for each year.
The values of $T_r$ are shown in Fig.\ \ref{Fig:income}.  The plots
for the 1980s and 1990s are shifted vertically for clarity.  We
observe that the data points in the lower-income part of the
distribution collapse on the same exponential curve for all years.
This demonstrates that the shape of the income distribution for the
lower class is extremely stable and does not change in time, despite
gradual increase in the average income in nominal dollars.  This
observation suggests that the lower-class distribution is in
statistical ``thermal'' equilibrium.

On the other hand, as Fig.\ \ref{Fig:income} shows, income
distribution of the upper class does not rescale and significantly
changes in time.  \textcite{Silva-2005} found that the exponent
$\alpha$ of the power law $C(r)\propto1/r^\alpha$ decreased from 1.8
in 1983 to 1.4 in 2000.  This means that the upper tail became
``fatter''.  Another useful parameter is the total income of the upper
class as the fraction $f$ of the total income in the system.  The
fraction $f$ increased from 4\% in 1983 to 20\% in 2000
\cite{Silva-2005}.  However, in year 2001, $\alpha$ increased and $f$
decreased, indicating that the upper tail was reduced after the stock
market crash at that time.  These results indicate that the upper tail
is highly dynamical and not stationary.  It tends to swell during the
stock market boom and shrink during the bust.  Similar results were
found for Japan
\cite{Souma-2001,Souma-2002,Fujiwara-2003,Aoyama-2003}.

\begin{figure}
\includegraphics[angle=-90,width=0.78\linewidth]{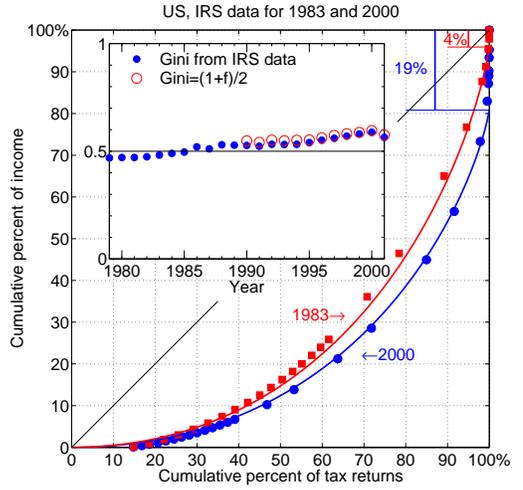}
\caption{\textit{Main panel:} Lorenz plots for income distribution in
  1983 and 2000.  The data points are from the IRS
  \cite{Strudler-2003}, and the theoretical curves represent
  Eq.\ (\ref{Lorenz}) with the parameter $f$ deduced from
  Fig.\ \ref{Fig:income}.  \textit{Inset:} The closed circles are the
  IRS data \cite{Strudler-2003} for the Gini coefficient $G$, and the
  open circles show the theoretical formula $G=(1+f)/2$.  From 
  \textcite{Silva-2005}.}
\label{Fig:Lorenz}
\end{figure}

Although relative income inequality within the lower class remains
stable, the overall income inequality in the USA has increased
significantly as a result of the tremendous growth of the income of
the upper class.  This is illustrated by the Lorenz curve and the Gini
coefficient shown in Fig.\ \ref{Fig:Lorenz}.  The Lorenz curve
\cite{Kakwani-book} is a standard way of representing income
distribution in the economic literature.  It is defined in terms of
two coordinates $x(r)$ and $y(r)$ depending on a parameter $r$:
\begin{equation}
  x(r)=\int_0^r P(r')\,dr',\quad
  y(r)=\frac{\int_0^r r' P(r')\,dr'}{\int_0^\infty r' P(r')\,dr'}.
\label{xy}
\end{equation}
The horizontal coordinate $x(r)$ is the fraction of the population
with income below $r$, and the vertical coordinate $y(r)$ is the
fraction of the income this population accounts for.  As $r$ changes
from 0 to $\infty$, $x$ and $y$ change from 0 to 1 and parametrically
define a curve in the $(x,y)$ plane.

Fig.\ \ref{Fig:Lorenz} shows the data points for the Lorenz curves in
1983 and 2000, as computed by the IRS \cite{Strudler-2003}.
\textcite{Dragulescu-2001a} analytically derived the Lorenz curve
formula $y=x+(1-x)\ln(1-x)$ for a purely exponential distribution
$P(r)=c\exp(-r/T_r)$.  This formula is shown by the upper curve in
Fig.\ \ref{Fig:Lorenz} and describes the 1983 data reasonably well.
However, for year 2000, it is essential to take into account the
fraction $f$ of income in the upper tail, which modifies for the
Lorenz formula as follows
\cite{Dragulescu-2003,Yakovenko-2005,Silva-2005}
\begin{equation}
  y=(1-f)[x+(1-x)\ln(1-x)]+f\,\Theta(x-1).
\label{Lorenz}
\end{equation}
The last term in Eq.\ (\ref{Lorenz}) represent the vertical jump of
the Lorenz curve at $x=1$, where a small percentage of population
in the upper class accounts for a substantial fraction $f$ of the
total income.  The lower curve in Fig.\ \ref{Fig:Lorenz} shows that
Eq.\ (\ref{Lorenz}) fits the 2000 data very well.

\begin{figure}
\includegraphics[width=0.8\linewidth]{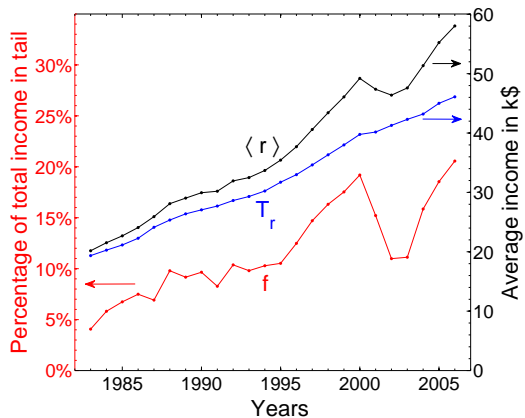}
\caption{Historical evolution of the parameters $\langle r\rangle$,
  $T_r$, and the fraction of income $f$ going to the upper tail, as
  defined in Eq.\ (\ref{f}).  From
  \textcite{Banerjee-2008}.}
\label{Fig:f}
\end{figure}

The deviation of the Lorenz curve from the straight diagonal line in
Fig.\ \ref{Fig:Lorenz} is a certain measure of income inequality.
Indeed, if everybody had the same income, the Lorenz curve would be
the diagonal line, because the fraction of income would be
proportional to the fraction of the population.  The standard measure
of income inequality is the Gini coefficient $0\leq G\leq1$, which is
defined as the area between the Lorenz curve and the diagonal line,
divided by the area of the triangle beneath the diagonal line
\cite{Kakwani-book}.  Time evolution of the Gini coefficient, as
computed by the IRS \cite{Strudler-2003}, is shown in the inset of
Fig.\ \ref{Fig:Lorenz}.  \textcite{Dragulescu-2001a} derived
analytically the result that $G=1/2$ for a purely exponential
distribution.  In the first approximation, the values of $G$ shown in
the inset of Fig.\ \ref{Fig:Lorenz} are indeed close to the
theoretical value 1/2.  If we take into account the upper tail using
Eq.\ (\ref{Lorenz}), the formula for the Gini coefficient becomes
$G=(1+f)/2$ \cite{Silva-2005}.  The inset in Fig.\ \ref{Fig:Lorenz}
shows that this formula gives a very good fit to the IRS data for the
1990s using the values of $f$ deduced from Fig.\ \ref{Fig:income}.
The values $G<1/2$ in the 1980s cannot be captured by this formula,
because the Lorenz data points are slightly above the theoretical
curve for 1983 in Fig.\ \ref{Fig:Lorenz}.  Overall, we observe that
income inequality has been increasing for the last 20 years, because
of swelling of the Pareto tail, but decreased in 2001 after the stock
market crash.

It is easy to show that the parameter $f$ in Eq.\ (\ref{Lorenz}) and
in Fig.\ \ref{Fig:Lorenz} is given by
\begin{equation}
  f=\frac{\langle r\rangle-T_r}{\langle r\rangle},
\label{f}
\end{equation}
where $\langle r\rangle$ is the average income of the whole
population, and the temperature $T_r$ is the average income in the
exponential part of the distribution.  Eq.\ (\ref{f}) gives a
well-defined measure of the deviation of the actual income
distribution from the exponential one and, thus, of the fatness of
the upper tail.  Fig.\ \ref{Fig:f} shows historical evolution of the
parameters $\langle r\rangle$, $T_r$, and $f$ given by
Eq.\ (\ref{f}).\footnote{A similar plot was constructed by
  \textcite{Silva-2005} for an earlier historical dataset.}  We
observe that $T_r$ has been increasing, essentially, monotonously (most
of this increase is inflation). In contrast, $\langle r\rangle$ had
sharp peaks in 2000 and 2006 coinciding with the speculative bubbles
in financial markets.  The fraction $f$, which characterizes income
inequality, has been increasing for the last 20 years and reached
maxima of 20\% in the years 2000 and 2006 with a sharp drop in
between.  We conclude that the speculative bubbles greatly increase
the fraction of income going to the upper tail, but do not change
income distribution of the lower class.  When the bubbles inevitably
collapse, income inequality reduces.

\begin{figure}
\includegraphics[width=0.9\linewidth]{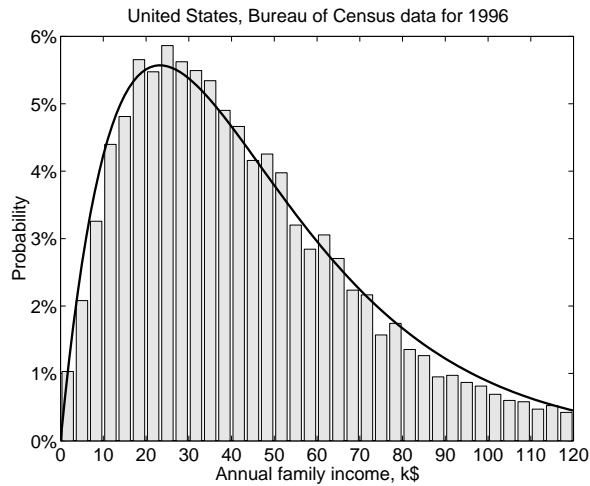}
\caption{\textit{Histogram:} Probability distribution of family income
  for families with two adults (US Census Bureau data).  \textit{Solid
  line:} Fit to Eq.\ (\ref{family}).  From
  \textcite{Dragulescu-2001a}.}
\label{Fig:income-2}
\end{figure}

\begin{figure}
\includegraphics[width=0.8\linewidth]{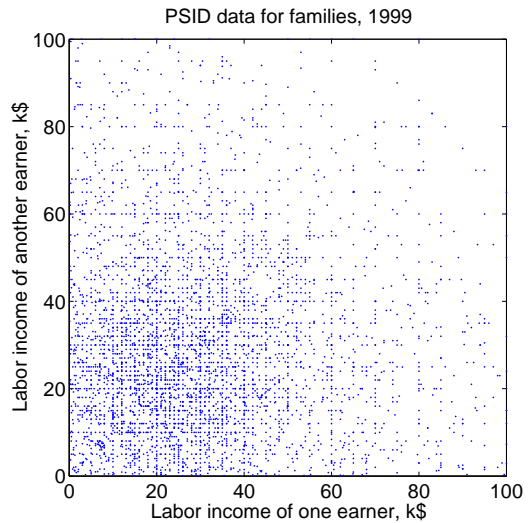}
\caption{Scatter plot of the spouses' incomes $(r_1,r_2)$ and
  $(r_2,r_1)$ based on the data from the Panel Study of Income
  Dynamics (PSID).   From \textcite{Dragulescu-2003}.}
\label{Fig:scatter}
\end{figure}

Thus far we discussed the distribution of individual income.  An
interesting related question is the distribution $P_2(r)$ of family
income $r=r_1+r_2$, where $r_1$ and $r_2$ are the incomes of spouses.
If the individual incomes are distributed exponentially
$P(r)\propto\exp(-r/T_r)$, then
\begin{equation}
  P_2(r)=\int_0^r dr' P(r')P(r-r')=c\,r\exp(-r/T_r),
\label{family}
\end{equation}
where $c$ is a normalization constant.  Fig.\ \ref{Fig:income-2} shows
that Eq.\ (\ref{family}) is in good agreement with the family income
distribution data from the US Census Bureau \cite{Dragulescu-2001a}.
In Eq.\ (\ref{family}), we assumed that incomes of spouses are
uncorrelated.  This simple approximation is indeed supported by the
scatter plot of incomes of spouses shown in Fig.\ \ref{Fig:scatter}.
Each family is represented in this plot by two points $(r_1,r_2)$ and
$(r_2,r_1)$ for symmetry.  We observe that the density of points is
approximately constant along the lines of constant family income
$r_1+r_2=\rm const$, which indicates that incomes of spouses are
approximately uncorrelated.  There is no significant clustering of
points along the diagonal $r_1=r_2$, i.e.,\ no strong positive
correlation of spouses' incomes.

\begin{figure}
\includegraphics[width=0.78\linewidth]{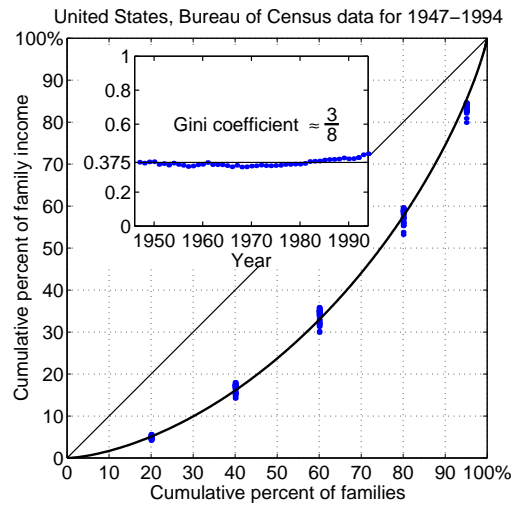}
\caption{\textit{Main panel:} Lorenz plot for family income calculated
  from Eq.\ (\ref{family}), compared with the US Census data points.
  \textit{Inset:} The US Census data points for the Gini coefficient
  for families, compared with the theoretically calculated value
  3/8=37.5\%.  From \textcite{Dragulescu-2001a}.}
\label{Fig:Lorenz-2}
\end{figure}

The Gini coefficient for the family income distribution (\ref{family})
was analytically calculated by \textcite{Dragulescu-2001a} as
$G=3/8=37.5\%$.  Fig.\ \ref{Fig:Lorenz-2} shows the Lorenz quintiles
and the Gini coefficient for 1947--1994 plotted from the US Census
Bureau data \cite{Dragulescu-2001a}.  The solid line, representing the
Lorenz curve calculated from Eq.\ (\ref{family}), is in good agreement
with the data.  The systematic deviation for the top 5\% of earners
results from the upper tail, which has a less pronounced effect on
family income than on individual income, because of income averaging
in the family.  The Gini coefficient, shown in the inset of
Fig.\ \ref{Fig:Lorenz-2}, is close to the calculated value of
$37.5\%$.  Notice that income distribution is very stable for a long
period of time, which was also recognized by economists
\cite{Levy-1987}.  Moreover, the average $G$ for the developed
capitalist countries of North America and western Europe, as
determined by the World Bank, is also close to the calculated value
37.5\% \cite{Dragulescu-2003}.  However, within this average, nations
or groups of nations may have quite different Gini coefficients that
persist over time due to specific historical, political, or social
circumstances \cite{Rosser-2004}.  The Nordic economies, with their
famously redistributive welfare states, have $G$ in the
mid-20\%, while many of the Latin American countries have $G$ over
50\%, reflecting entrenched social patterns inherited from the
colonial era.

Income distribution has been examined in econophysics papers for
different countries: Japan
\cite{Souma-2001,Souma-2002,Fujiwara-2003,Aoyama-2003,Souma-2005,Nirei-2007,Ferrero-2004,Ferrero-2005},
Germany \cite{Clementi-2005a,Clementi-2007}, the UK
\cite{Richmond-2006b,Ferrero-2004,Ferrero-2005,Clementi-2005a,Clementi-2007},
Italy \cite{Clementi-2005b,Clementi-2006,Clementi-2007}, the USA
\cite{Clementi-2005a,Rawlings-2004}, India \cite{Sinha-2006},
Australia \cite{DiMatteo-2004,Clementi-2006,Banerjee-2006}, and New
Zealand \cite{Ferrero-2004,Ferrero-2005}.  The distributions are
qualitatively similar to the results presented in this section.  The
upper tail follows a power law and comprises a small fraction of
population.  To fit the lower part of the distribution, different
papers used the exponential, Gamma, and log-normal distributions.
Unfortunately, income distribution is often reported by statistical
agencies for households, so it is difficult to differentiate between
one-earner and two-earner income distributions.  Some papers used
interpolating functions with different asymptotic behavior for low and
high incomes, such as the Tsallis function \cite{Ferrero-2005} and the
Kaniadakis function \cite{Clementi-2007}.  However, the transition
between the lower and upper classes is not smooth for the US data
shown in Figs.\ \ref{Fig:income1997} and \ref{Fig:income}, so such
functions would not be useful in this case.  The special case is
income distribution in Argentina during the economic crisis, which
shows a time-dependent bimodal shape with two peaks
\cite{Ferrero-2005}.

\subsection{Theoretical models of income distribution}
\label{Sec:r-theory}

Having examined the empirical data on income distribution, let us now
discuss theoretical models.  Income $r_i$ is the influx of money per
unit time to an agent $i$.  If the money balance $m_i$ is analogous to
energy, then the income $r_i$ would be analogous to power, which is
the energy flux per unit time.  So, one should conceptually
distinguish between the distributions of money and income.  While
money is regularly transferred from one agent to another in pairwise
transactions, it is not typical for agents to trade portions of their
income.  Nevertheless, indirect transfer of income may occur when one
employee is promoted and another demoted while the total annual budget
is fixed, or when one company gets a contract whereas another one
loses it, etc.  A reasonable approach, which has a long tradition in
the economic literature
\cite{Gibrat-1931,Kalecki-1945,Champernowne-1953}, is to treat
individual income $r$ as a stochastic process and study its
probability distribution.  In general, one can study a Markov process
generated by a matrix of transitions from one income to another.  In
the case where the income $r$ changes by a small amount $\Delta r$
over a time period $\Delta t$, the Markov process can be treated as
income diffusion.  Then one can apply the general Fokker-Planck
equation \cite{Kinetics-book} to describe evolution in time $t$ of the
income distribution function $P(r,t)$ \cite{Silva-2005}
\begin{equation}
   \frac{\partial P}{\partial t}=\frac{\partial}{\partial r} \left[AP
   + \frac{\partial(BP)}{\partial r}\right], \; A=-{\langle\Delta
   r\rangle \over \Delta t}, \; B={\langle(\Delta r)^2\rangle \over
   2\Delta t}.
\label{diffusion}
\end{equation}
The coefficients $A$ and $B$ in Eq.\ (\ref{diffusion}) are determined
by the first and second moments of income changes per unit time.  The
stationary solution $\partial_tP=0$ of Eq.\ (\ref{diffusion}) obeys
the following equation with the general solution
\begin{equation}
  \frac{\partial(BP)}{\partial r}=-AP,\quad
  P(r)=\frac{c}{B(r)}\exp\left(-\int^r\frac{A(r')}{B(r')}dr'\right).
\label{stationary}
\end{equation}

For the lower part of the distribution, it is reasonable to assume
that $\Delta r$ is independent of $r$, i.e.,\ the changes in income
are independent of income itself.  This process is called the additive
diffusion \cite{Silva-2005}.  In this case, the coefficients in
Eq.\ (\ref{diffusion}) are the constants $A_0$ and $B_0$.  Then
Eq.\ (\ref{stationary}) gives the exponential distribution
$P(r)\propto\exp(-r/T_r)$ with the effective income temperature
$T_r=B_0/A_0$.\footnote{Notice that a meaningful stationary solution
  (\ref{stationary}) requires that $A>0$, i.e.,\ $\langle\Delta
  r\rangle<0$.}  The coincidence of this result with the
Boltzmann-Gibbs exponential law (\ref{Gibbs}) and (\ref{money}) is not
accidental.  Indeed, instead of considering pairwise interaction
between particles, one can derive Eq.\ (\ref{Gibbs}) by considering
energy transfers between a particle and a big reservoir, as long as
the transfer process is ``additive'' and does not involve a
Maxwell-demon-like discrimination \cite{Basu-2008}.  Although
money and income are different concepts, they may have similar
distributions, because they are governed by similar mathematical
principles.  It was shown explicitly by
\textcite{Dragulescu-2000,Slanina-2004,Cordier-2005} that the models
of pairwise money transfer can be described in a certain limit by the
Fokker-Planck equation.

On the other hand, for the upper tail of income distribution, it is
reasonable to expect that $\Delta r\propto r$, i.e.,\ income changes
are proportional to income itself.  This is known as the
proportionality principle of \textcite{Gibrat-1931}, and the process
is called the multiplicative diffusion \cite{Silva-2005}.  In this
case, $A=ar$ and $B=br^2$, and Eq.\ (\ref{stationary}) gives the
power-law distribution $P(r)\propto1/r^{\alpha+1}$ with
$\alpha=1+a/b$.

\begin{figure}
\includegraphics[width=0.7\linewidth,clip]{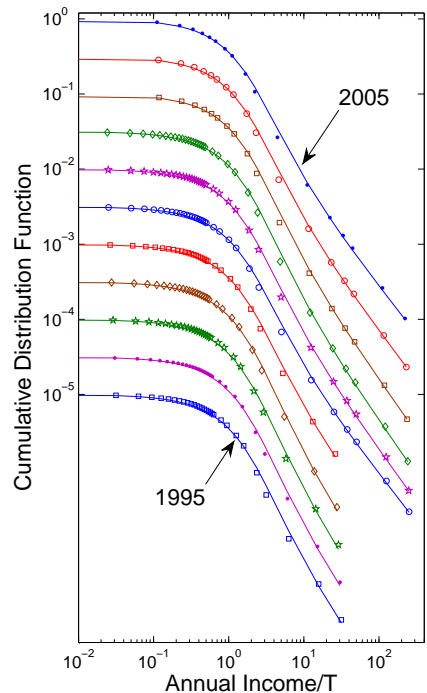}
\caption{Fits of the IRS data for income distribution using
  Eq.\ (\ref{arctan}).  Plots for different years are shifted
  vertically for clarity.  From \textcite{Banerjee-2008}.}
\label{Fig:+x}
\end{figure}

Generally, the lower-class income comes from wages and salaries, where
the additive process is appropriate, whereas the upper-class income
comes from bonuses, investments, and capital gains, calculated in
percentages, where the multiplicative process applies
\cite{Milakovic-2005}.  However, the additive and multiplicative
processes may coexist.  An employee may receive a cost-of-living raise
calculated in percentages (the multiplicative process) and a merit
raise calculated in dollars (the additive process).  Assuming that
these processes are uncorrelated, we have $A=A_0+ar$ and
$B=B_0+br^2=b(r_0^2+r^2)$, where $r_0^2=B_0/b$.  Substituting these
expressions into Eq.\ (\ref{stationary}), we find
\begin{equation}
  P(r)=c\,\frac{e^{-(r_0/T_r)\arctan(r/r_0)}}
  {[1+(r/r_0)^2]^{1+a/2b}}.
\label{arctan}
\end{equation}
The distribution (\ref{arctan}) interpolates between the exponential
law for low $r$ and the power law for high $r$, because either the
additive or the multiplicative process dominates in the corresponding
limit.  The crossover between the two regimes takes place at $r=r_0$,
where the additive and multiplicative contributions to $B$ are equal.
The distribution (\ref{arctan}) has three parameters: the ``income
temperature'' $T_r=A_0/B_0$, the Pareto exponent $\alpha=1+a/b$, and
the crossover income $r_0$.  It is a minimal model that captures the
salient features of the empirical income distribution.
Eq.\ (\ref{arctan}) was obtained by \textcite{Yakovenko-2009}, and a
more general formula for correlated additive and multiplicative
processes was derived by \textcite{Fiaschi-Marsili} for a
sophisticated economic model.  Fits of the IRS data using
Eq.\ (\ref{arctan}) are shown in Fig.\ \ref{Fig:+x} reproduced from
\textcite{Banerjee-2008}.  A mathematically similar, but more
economically oriented, model was proposed by
\textcite{Souma-2005,Nirei-2007}, where labor income and assets
accumulation are described by the additive and multiplicative
processes correspondingly.  A general stochastic process with additive
and multiplicative noise was studied numerically by
\textcite{Takayasu-1997}, but the stationary distribution was not
derived analytically.  A similar process with discrete time increments
was studied by \textcite{Kesten-1973}.  Besides economic applications,
Eq.\ (\ref{arctan}) may be also useful for general stochastic
processes with additive and multiplicative components.

To verify the multiplicative and additive hypotheses empirically, it
is necessary to have data on income mobility, i.e.,\ the income
changes $\Delta r$ of the same people from one year to another.  The
distribution of income changes $P(\Delta r|r)$ conditional on income
$r$ is generally not available publicly, although it can be
reconstructed by researchers at the tax agencies.  Nevertheless, the
multiplicative hypothesis for the upper class was quantitatively
verified by \textcite{Fujiwara-2003,Aoyama-2003} for Japan, where such
data for the top taxpayers are publicly available.

The power-law distribution is meaningful only when it is limited to
high enough incomes $r>r_0$.  If all incomes $r$ from 0 to $\infty$
follow a purely multiplicative process ($A_0=0$ and $B_0=0$), then one
can change to a logarithmic variable $x=\ln(r/r_*)$ in
Eq.\ (\ref{diffusion}) and show that it gives a Gaussian
time-dependent distribution $P_t(x)\propto\exp(-x^2/2\sigma^2t)$ for
$x$, i.e.,\ the log-normal distribution for $r$, also known as the
Gibrat distribution \cite{Gibrat-1931}.  However, the width of this
distribution increases in time, so the distribution is not
stationary.  This was pointed out by \textcite{Kalecki-1945} a long
time ago, but the log-normal distribution is still widely used for
fitting income distribution, despite this fundamental logical flaw in
its justification.  In the classic paper, \textcite{Champernowne-1953}
showed that a multiplicative process gives a stationary power-law
distribution when a boundary condition is imposed at $r_0\neq0$.
Later, this result was rediscovered by econophysicists
\cite{Levy-1996,Sornette-1997,Levy-2003}.  In Eq.\ (\ref{arctan}),
the exponential distribution of the lower class effectively provides
such a boundary condition for the power law of the upper class.
Notice also that Eq.\ (\ref{arctan}) reduces to
Eq.\ (\ref{P-Bouchaud}) in the limit $r_0\to0$ with $B_0=0$, but
$A_0\neq0$.

There are alternative approaches to income distribution in economic
literature.  One of them, proposed by \textcite{Lydall-1959}, involves
social hierarchy.  Groups of people have leaders, which have leaders
of the higher order, and so on.  The number of people decreases
geometrically (exponentially) with the increase in the hierarchical
level. If individual income increases by a certain factor
(i.e.,\ multiplicatively) when moving to the next hierarchical level,
then income distribution follows a power law \cite{Lydall-1959}.
However, this original argument of Lydall can be easily modified to
produce the exponential distribution.  If individual income increases
by a certain amount, i.e.,\ income increases linearly with the
hierarchical level, then income distribution is exponential.  The
latter process seems to be more realistic for moderate annual incomes
below 100~k\$.  A similar scenario is the Bernoulli trials
\cite{Feller-book}, where individuals have a constant probability of
increasing their income by a fixed amount.  We see that the
deterministic hierarchical models and the stochastic models of
additive and multiplicative income mobility represent essentially the
same ideas.

\section{Conclusions}
\label{Sec:conclusions}

The ``invasion'' of physicists into economics and finance at the turn
of the millennium is a fascinating phenomenon.  It generated a lively
public debate about the role and future perspectives of econophysics,
covering both theoretical and empirical issues.\footnote{See, for
  example,
  \textcite{Lux-2005,Anglin-2005,Gallegati-2006,Lux-2008,McCauley-2006,response-2006,Rosser-2008b,Rosser-2006,Ball-2006,Stauffer-history,Keen-2008,Yakovenko-2009,Carbone-2007}.}
The econophysicist Joseph McCauley proclaimed that ``Econophysics will
displace economics in both the universities and boardrooms, simply
because what is taught in economics classes doesn't work''
\cite{Ball-2006}.  Although there is some truth in his arguments
\cite{McCauley-2006}, one may consider a less radical scenario.
Econophysics may become a branch of economics, in the same way as game
theory, psychological economics, and now agent-based modeling became
branches of economics.  These branches have their own interests,
methods, philosophy, and journals.  When infusion of new ideas from a
different field happens, the main contribution often consists not in
answering old questions, but in raising new questions.  Much of the
misunderstanding between economists and physicists happens not because
they are getting different answers, but because they are answering
different questions.

The subject of income and wealth distributions and social inequality
was very popular at the turn of another century and is associated with
the names of Pareto, Lorenz, Gini, Gibrat, and Champernowne, among
others.  Following the work by Pareto, attention of researchers was
primarily focused on the power laws.  However, when physicists took a
fresh look at the empirical data, they found a different, exponential
law for the lower part of the distribution.  Demonstration of the
ubiquitous nature of the exponential distribution for money, wealth,
and income is one of the new contributions produced by
econophysics.\footnote{The exponential distribution is also ubiquitous
  in the probability distributions of financial returns
  \cite{Silva-2004,Kleinert-2007,McCauley-2003} and the growth rates
  of firms.}  The motivation, of course, came from the Boltzmann-Gibbs
distribution in physics. Further studies revealed a more detailed
picture of the two-class distribution in a society.  Although social
classes have been known in political economy since Karl Marx,
realization that they are described by simple mathematical
distributions is quite new.  Very interesting work was done by the
computer scientist Ian \textcite{Wright-2005,Wright-2009}, who
demonstrated emergence of two classes in an agent-based simulation of
initially equal agents.  This work has been further developed in the
upcoming book by \textcite{Cottrell-2009}, integrating economics,
computer science, and physics.

Econophysics may be also useful for teaching of statistical physics.
If nothing else, it helps to clarify the foundations of statistical
physics by applying it to nontraditional objects.  Practitioners of
statistical physics know very well that the major fascinating
attraction of this field is the enormous breadth of its applications.


\end{document}